%% file: main.tex
\documentclass[journal]{IEEEtran}

\usepackage{fancyhdr}
\input{packages}

\begin{document}

\title{\acronym: Leveraging Composability and Diversity to Design Fault and Intrusion Resilient Chips}

\author{Ahmad~T.~Sheikh$^\dag$,~\IEEEmembership{Member,~IEEE,}
Ali~Shoker, 
Suhaib~A.~Fahmy,~\IEEEmembership{Senior Member,~IEEE,}
and~Paulo~Esteves-Verissimo,~\IEEEmembership{Fellow,~IEEE}
\thanks{The authors are with the CEMSE Division, King Abdullah University of Science and Technology (KAUST)
Thuwal 23955-6900, Kingdom of Saudi Arabia. (emails: \{ahmad.sheikh, ali.shoker, suhaib.fahmy, paulo.verissimo\{@kaust.edu.sa\}\})}
\thanks{$^\dag$Correspondence address: ahmad.sheikh@kaust.edu.sa}}

\maketitle
\thispagestyle{fancy}


\input{abstract}

\input{introduction}

\input{related-works}

\input{attack-model}

\input{diversity-for-resilience}

\input{egraph}
\input{resilogic-framework}

\input{results}

\input{conclusion}

\bibliographystyle{IEEEtran}
\bibliography{bib-paper}


\end{document}

%% file: packages.tex
\usepackage{algorithm}
\usepackage{algpseudocode}
\usepackage{array}
\usepackage{url}
\usepackage{hyperref}
\usepackage{cite}
\usepackage{amsmath,amssymb,amsfonts}
\usepackage{graphicx}
\usepackage{textcomp}
\usepackage{xcolor}
\usepackage{subcaption}
\usepackage{multirow}
\usepackage{blkarray}
\usepackage[english]{babel}
\usepackage{amsthm}
\usepackage{xspace}
\usepackage{verbatim}
\usepackage{textcomp}
\usepackage{stfloats}
\usepackage{balance}

\newcommand\acronym{\textit{ResiLogic}\xspace}

\newlength\myindent
\newcommand{\rulesep}{\unskip\ \vrule\ }

\newtheorem{definition}{Definition}[section]    
\newtheorem{theorem}{Theorem}[section]
\newtheorem{lemma}[theorem]{Lemma}


%% file: abstract.tex
\begin{abstract}
A long-standing challenge is the design of chips resilient to faults and glitches. Both fine-grained gate diversity and coarse-grained modular redundancy have been used in the past. However, these approaches have not been well-studied under other threat models where some stakeholders in the supply chain are untrusted. Increasing digital sovereignty tensions raise concerns regarding the use of foreign off-the-shelf tools and IPs, or off-sourcing fabrication, driving research into the design of resilient chips under this threat model. This paper addresses a threat model considering three pertinent attacks to resilience: distribution, zonal, and compound attacks. To mitigate these attacks, we introduce the \acronym framework that exploits \textit{Diversity by Composability}: constructing diverse circuits composed of smaller diverse ones by design. This approach enables designers to develop circuits in the early stages of design without the need for additional redundancy in terms of space or cost. To generate diverse circuits, we propose a technique using E-Graphs with new rewrite definitions for diversity. Using this approach at different levels of granularity is shown to improve the resilience of circuit design in \acronym up to $\times 5$ against the three considered attacks.
\end{abstract}

\begin{IEEEkeywords}
Chip Resilience, Hardware Diversity, TMR, Fault and Intrusion Tolerance (FIT), E-Graphs 
\end{IEEEkeywords}

%% file: introduction.tex
\section{Introduction}
\label{sec:introduction}
\IEEEPARstart{I}{ntegrated} Circuit (IC) development is a pressing challenge in the modern digital ecosystem. The huge supply chain and toolchain across the development pipeline involves a large number of \textit{known and unknown} stakeholders and dependencies. This has compounded the development complexity and results in higher likelihood of \textit{unintentional faults}, glitches, bugs, etc.~\cite{bar2006sorcerer, Mazumder:2023/1769} and other \textit{intentional intrusions} with malicious intent~\cite{farimah2020system}. While resilience to unintentional benign faults has been well studied and mitigated in the literature, mainly using hardening and redundancy~\cite{verissimo2007intrusion, sheikh2018double, hsu1992time, townsend2003quadruple, bas2022safede, tambara2013evaluating}, the case of resilience against malicious actors (e.g., vendors and foundries) has often been ignored. 
With nation states aligning their priorities to develop local semiconductor manufacturing, technological excellence, and bridge recent global chip shortages~\cite{us_chip_act-2022, eu_chip_act-2023, chips-shortage}, the quest for new resilience methods to withstand untrusted stakeholders is regaining momentum.

\begin{figure}[ht]
\centering
    \includegraphics[width=0.8\linewidth]{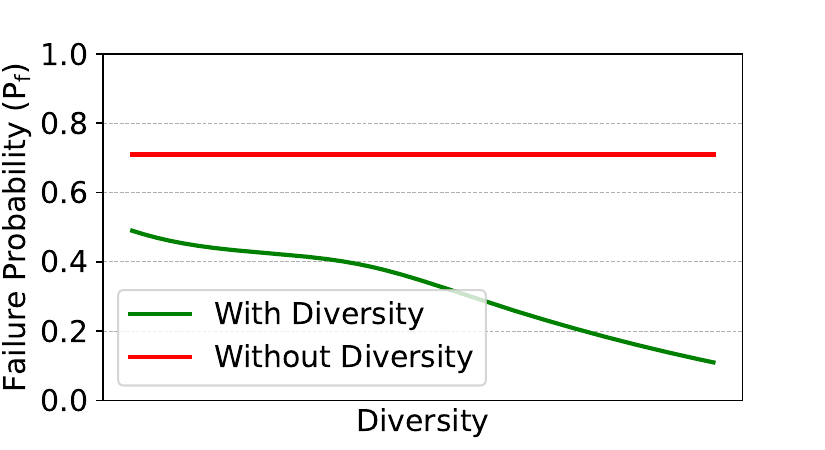}
    \caption{Multi-level diversity (used in \acronym) has a significant impact on tolerating common mode failures.}
    \label{fig:diversity-impact}
\end{figure}%

Our work is the first that studies chip design resiliency assuming a threat model in which vendors and silicon fabs may be untrusted.
Typical chip design can be more productive when the designer relies on available off-the-shelf tools and libraries from different vendors and then sends the design to a silicon fab for manufacturing. The premise is that the vendors and fabs are trusted, both ethically and technically. With the recent digital sovereignty tensions, the former is not guaranteed and hardly measurable, while the latter has been shown to be unrealistic with such a complex development process, incurring too many dependencies on third party tools, libraries, and sometime open source resources~\cite{ajayi2019openroad}. Our analysis shows that a design is highly prone to three potential attacks of relevance to resiliency (as outlined in Section~\ref{sec:attack-model}): \textit{Distribution} attacks (supply chain trojans and backdoors), \textit{Zonal} attacks (using electromagnetic or \textit{Laser} beam interference), and \textit{Compound} attacks (simultaneous \textit{Distribution} and \textit{Zonal} attacks).

Redundancy in space is a fundamental resilience approach that is used to mask faults and intrusions~\cite{inoue2009gmr,verissimo2007intrusion}. 
Two main chip resiliency directions for redundancy at different granularity are studied in state of the art chip design. The first is fine-grained, where redundancy is applied at the logic gate level~\cite{afzaal2022evolutionary}. This approach aims at diversifying the logic circuit design through adding various basic logic gates, e.g., AND and Inverter. While this approach has been shown to be effective, e.g., increasing design resilience up to 90\% under (benign) faults~\cite{miller2008cartesian}, it is designed to be incorporated in the vendor's IP libraries and tools, where the designer has little control. The designer must either  create libraries (costly and unproductive) or live under the mercy of the vendor and fab (who may collude).

To facilitate diversification, we provide a new technique to generate diverse circuits using E-Graphs~\cite{2021-egg}. While E-Graphs have been often used to optimize circuits, we introduce new equality saturation rewrite definitions to produce functionally equivalent but structurally different artifacts at a reasonable overhead for several benchmarks.

The second redundancy approach is course-gained, following the \textit{Triple Modular Redundancy}~\cite{anghel2000evaluation,sheikh2018double} (TMR) style replication that requires three identical copies of a circuit followed by a majority voter to mask the faulty one. However, this works well as long as a minority of circuits are non-faulty and assuming the independence of failures, i.e., the absence of \textit{Common Mode Failures} (CMF, henceforth); this is often considered a hard assumption~\cite{avizienis1995methodology, garcia2014analysis}. In addition, this often incurs high replication cost (i.e., often a factor of three). As shown in Fig.~\ref{fig:diversity-impact}, the average failure probability of TMR  against CMF (red curve) is around 70\%, even when used together with the aforementioned E-Graph approach~\cite{miller2008cartesian} (see more details in Section~\ref{sec:results-discussions}). 

We address this by enabling the concept of \textit{composability} that leverages the properties of higher level composable logic circuits, i.e., circuits whose modular structure is composed of smaller linked circuits. Together with diversification, composability boosts the resilience without incurring extra replication costs, thanks to the implicit modularity of composable circuits.
This holds for various circuits that are composable by nature, e.g., \textit{Systolic Array}, \textit{N-bit Multiplier}, \textit{N-bit Adder}, \textit{Vector Unit}, etc.~\cite{koren2018computer, jouppi2018motivation}, which makes the approach feasible for several modern applications (see Section~\ref{sec:diversity-for-resilience} for details).

Our third contribution is introducing \acronym, a framework for building resilient diversified integrated circuits at a reasonable overhead. 
\acronym allows for designing resilient circuits through employing different levels of redundancy and diversity combining diversification and composability. \acronym provide specifications and tools to cover the process from diverse circuit generation, CMF testing, all the way to different levels of redundancy (by composability and replication). This results in a significant reduction in the probability of failures as shown in Fig.~\ref{fig:diversity-impact}. Our evaluation demonstrates that the circuit resilience, like \textit{N-bit Adder} can withstand Distribution, Zonal, and Compound attacks up to a factor of five.

The rest of the paper is organized as follows: related work is discussed in~\ref{sec:related-works},  Section~\ref{sec:attack-model} introduces the threat and fault models. Section~\ref{sec:diversity-for-resilience} introduces the notion of \textit{Diversity by Composability} that is the basis for the \acronym framework discussed in Section~\ref{sec:resilogic-framework}. 
\acronym evaluation strategy and results are presented in Section~\ref{sec:results-discussions}, followed by the concluding remarks in Section~\ref{sec:conclusion}, respectively. 

%% file: related-works.tex
\section{Related Work}
\label{sec:related-works}
The resilience of a system is significantly improved by applying replication followed by a voter for majority consensus. Traditional approaches to avoid CMF in TMR/DMR have been to implement diverse systems at higher levels of abstraction~\cite{borges2010diversity, tambara2013evaluating}. Faults can be intentional or accidental and have predictable manifestations in digital circuits.
State of the art literature classifies faults/vulnerabilities into four categories: 1) hardware trojans or malicious implants, 2) backdoor through test/debug interfaces, 3) accidental/unintentional vulnerabilities~\cite{farimah2020system, Mazumder:2023/1769}, and 4) external faults~\cite{karaklajic2013hardware, fiolhais2023transient}.
To combat low level faults, resilience through design diversity must be applied at the gate level.
Cartesian Genetic Programming (CGP)~\cite{miller2008cartesian} had been proposed to generate diverse isofunctional structures to improve resilience at the gate level of a design~\cite{afzaal2022evolutionary}. Similarly, redundant designs can be spatially separated to provide resilience against external faults in TMR systems~\cite{tmr-spatial}. \acronym attempts to build on this concept of diversity, however, proposes a novel approach to combine small diverse structures to create larger diverse artifacts.

Faults based on external attacks can be due to various sources.
Power supplies can be tampered with to alter uniform behavior, resulting in erroneous results.
Similarly, power spiking can cause a processor to not only misinterpret an instruction but also induce memory faults~\cite{balasch2011depth}. 
Clock glitches, through a deviated clock signal, can be induced to cause execution of subsequent instructions before retirement of previous instructions, i.e., modification of the program counter (PC) at irregular intervals~\cite{bar2006sorcerer}.
This type of attack is considered to be the simplest as low-end FPGAs can be used to inject clock glitches~\cite{balasch2011depth, endo2011chip}. 
Electronic devices function properly in a certain temperature range, however, when exposed to extreme temperatures, the data inside memories can be corrupted.
These attacks are easier to set up but their effect cannot be localized.
Optical attacks are performed by exposing the device to a focused laser beam or strong photo flash.
This attack  can be targeted on either side of the chip and can be confined to certain area to set or reset memory bits or switch transistor state~\cite{asadizanjani2021optical}.
Similarly, electromagnetic (EM) interference can corrupt or modify memory contents by inducing eddy currents, resulting in a single or multiple bit faults~\cite{farahmandi2023cad}.
Both optical and electromagnetic attack fits our definition of zonal attacks in this paper. 

Attacks based on malicious implants can be inserted into a design at various stages, such as by an untrusted CAD designer, by a malicious tool, or at the foundry~\cite{bhunia2014hardware}.
Furthermore, hardware verification is a time-consuming and costly task, making it challenging to detect such trojans at scale~\cite{Yamashita2008, tehranipoor2010survey, shah2020impact}.
Malicious implants can be activated by an attacker using a carefully crafted strategy~\cite{adee2008hunt}.
Previous works have discussed hardware trojan activation using a combination of software-hardware~\cite{ancajas2014fort} and triggers based on time, input sequence and traffic patterns~\cite{bhunia2018hardware}.
We model these \textit{Distribution Attacks} by injecting fault(s) at specific location(s) in the design and investigate which input patterns are able to activate them. 


%% file: attack-model.tex
\section{Threat and Fault Models}
\label{sec:attack-model}

\subsection{Actors and Adversaries}
We consider a typical design process of three main actors in the ecosystem: \textit{Designer}, \textit{Vendor}, and \textit{Fab}. The designer may use several products from several vendors  (e.g., many tools and implementations) to develop a resilient chip, which is sent to the fab for fabrication. The roles and trust models of the actors are as follows:
\begin{itemize}
    \item \textit{Designer}: the designer of the chip and the main user of \acronym. They are assumed to be a trusted entity, and a target/victim of acting adversaries.
    \item \textit{Vendor}: the provider of pre-designed prerequisites like design libraries, implementations, or toolchains. Not all vendors are trusted entities.
    \item \textit{Fab}: the post-design fabrication foundry that uses silicon technology to implement the design as an integrated circuit. Not all fabs are trusted entities. 
\end{itemize}

\subsection{Intrusions and Faults}
\begin{figure}[t]
\centering
    \begin{subfigure}{0.15\textwidth}
        \includegraphics[width=\textwidth]{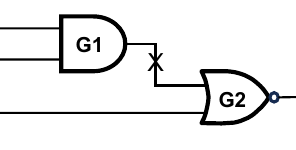}
        \caption{Injection}
        \label{fig:fault-inj}
    \end{subfigure}
    \hfill
    \begin{subfigure}{0.15\textwidth}
        \includegraphics[width=\textwidth]{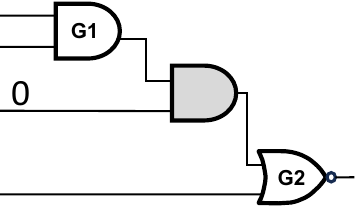}
        \caption{\texttt{stuck-at-0}.}
        \label{fig:sa0-fault}
    \end{subfigure}
    \hfill
    \begin{subfigure}{0.15\textwidth}
        \includegraphics[width=\textwidth]{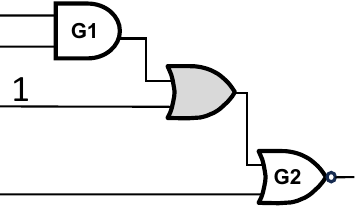}
        \caption{\texttt{stuck-at-1}.}
        \label{fig:sa1-fault}
    \end{subfigure}
    \caption{\texttt{stuck-at} intrusion/fault injection.}
    \label{fig:fault-model}
\end{figure}

During the design and manufacturing process, the design is prone to unintentional faults and intentional intrusions, leading to some functional circuit failure. Faults can be induced at any stage, even by the designer; while intrusions are assumed to be performed through attacking the circuit by the vendor or the fab, who can also collude. Intrusions and faults and their corresponding failures can manifest as \texttt{stuck-at} faults~\cite{mei1974bridging}.
In this model, the injected fault/intrusion on a logic gate results in a logic \texttt{0} (\texttt{stuck-at-0}) or \texttt{1} (\texttt{stuck-at-1}). Fig.~\ref{fig:fault-model} depicts a schematic of intrusion/fault injection between (a) two gates $G1$ and $G2$, leading to (b) \texttt{stuck-at-0} or (c) \texttt{stuck-at-1} faults. This can be replicated between any gates in the design.

\subsection{Threats and Attacks}
We consider the threat model where the goal of the adversaries, i.e., vendor and fab, is to break the \textit{integrity} of the design or circuit. The aim is to subvert the output either permanently or occasionally.  
We define three types of potential integrity attacks, i.e., Distribution, Zonal, and Compound attacks. SAT attacks~\cite{xie2016mitigating}, which are devised to extract the functionality of obfuscated circuits are not in the purview of this work.

\subsubsection{Distribution Attack}
It is typically faster and cheaper for the designer to use off-the-shelf tools, pre-designed modules, compilers, synthesizers, etc, to create a new design. Any malicious vendor in this chain can induce intrusions, trojans, or backdoors~\cite{farimah2020system}. 
The Distribution attack (or fault) is a supply-chain attack where a fault or intrusion can be induced at any step prior to or after design~\cite{xiao2016security,farimah2020system, Mazumder:2023/1769}. This attack targets any circuit module, channel, or zone. 



\subsubsection{Zonal Attack}
\begin{figure}[t]
\centering
        \includegraphics[width=0.5\columnwidth]{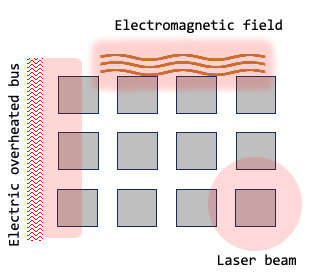}
        \caption{Possible zonal attacks.}
    \label{fig:zonal-attacks}
\end{figure}

The Zonal attack is a directed attack on the design layout. As shown in Fig.~\ref{fig:zonal-attacks}, the attack can target a chip in many ways, among them: (1) using an electric bus to overheat an edge~\cite{nagata2021physical}, (2) using an Electromagnetic field to subvert another edge~\cite{farahmandi2023cad,adee2008hunt}, or (3) using an optical laser beam to target a specific zone after fabrication~\cite{nagata2021physical}. 

\subsubsection{Compound Attack}
The Compound attack is a more complex attack where Distribution and Zonal attacks are preformed simultaneously. In this case, these attacks could be coordinated, e.g., collusion between vendor and fab, or uncoordinated. Although the former is more directed, both scenarios are devastating to the circuit as we show in the simulated experiments.

%% file: diversity-for-resilience.tex
\section{Diversity by Composability}
\label{sec:diversity-for-resilience}
In this section, we provide the motivations and design decisions behind \acronym.

\subsection{The Quest for Resilience}
Designing for resilience often requires redundancy in space or time~\cite{townsend2003quadruple}.
Redundancy in time entails repeating computation using the same hardware.
This is effective against transient failures, but not permanent faults and longer term intrusions.
Hence, redundancy in space through replicated hardware, with convergence to a final output is often used.
This final output can incorporate the replicated outputs or choose between them, depending on application.
Since most applications require determinism, reconciling outputs by majority voting is the most common approach in the literature~\cite{inoue2009gmr}.
This is usually on top of Double- or Triple-Modular-Redundancy (TMR). Majority voting only works if the replicated logic units do not fail together due to Common Mode Failures (CMF)~\cite{babu2019efficient}.
Consequently, modular redundancy exhibits very limited resilience (e.g, against transient failures) at a high hardware cost.
We enable the effectiveness of TMR through techniques to generate and use diverse replicas that are functionally equivalent but structurally different.

On the other hand, the replication cost of TMR may not be justified or affordable. This raises the challenge to find other ways to leverage diversity without including extra replication costs. In our work, we propose composability as an effective approach against untrusted supply chain stakeholders. Importantly, we show that composability preserves the determinism properties of the diversified logic circuits.  

\subsection{Diversity Levels}
Redundancy and diversity have been studied in the literature across the entire development process, involving all actors: vendors, designers, and fabs.
However, the threat model we discussed in Section~\ref{sec:attack-model} may render the redundancy implemented by a malicious vendor or unreliable fab.
This necessitates redundancy implemented by the designer to mitigate these potential malicious threats.
For integrated circuits, we identify three possible levels of diversity: Gate-level, Module-level, and Artifact-level diversity.
We are interested in providing the \textit{designer} with a mechanism to build resilient circuits, following a bottom-up approach, where diversity at the gate level is utilized to create larger diverse artifacts \textit{by composition}.

\subsubsection{\textbf{Gate-level diversity}}
\begin{figure}[t]
    \centering
        \begin{subfigure}{0.2\textwidth}
            \includegraphics[width=\textwidth]{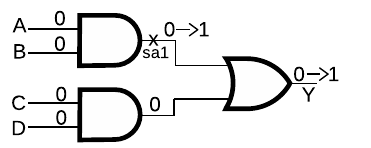}
            \caption{$Y=AB+CD$}
            \label{subfig:sop1}
       \end{subfigure}%
        \begin{subfigure}{0.2\textwidth}
            \includegraphics[width=\textwidth]{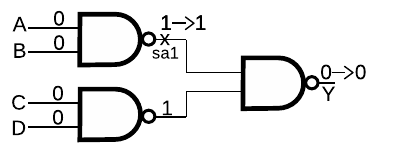}
            \caption{$Y=\overline{\overline{AB}\cdot\overline{CD}}$}
            \label{subfig:sop2}
        \end{subfigure}               
        \caption{Gate-level diversity highlighting different manifestations of the $stuck-at-1$ fault.}
        \label{fig:diverse-same-funcs}
\end{figure}

The lowest level of diversity we consider, targets designing diverse primitive logic modules composed of basic logic gates, e.g., AND, OR. 
The modules (e.g., adders, multipliers, accumulators, etc.) require resilience since they are core to the datapaths of larger designs, and importantly, they are often provided by IP vendors---which we assume to be untrusted.
Therefore, an intrusion or fault at this level can be replicated as much as these primitive logic units are used.
We require the designer to have access to diverse implementations preferably from different IP vendors or, if possible, constructed by the designer themselves.
This minimizes the likelihood of using diverse units from many malicious IP vendors at once, and thus reduces CMFs.
Diversity at this level can be achieved using different combinations of logic gates while maintaining the same functionality. This can be done manually, but preferably using an automated algorithm as we do in this work, inspired by~\cite{schafer2019high, alkabani2008n}. 
For instance, the equivalent boolean functions $AB+CD$ and $\overline{\overline{AB}\cdot\overline{CD}}$ will have different gate-level implementations that exhibit different behaviors under attacks (manifesting as \textit{stuck-at}). 
As an example, Fig.~\ref{fig:diverse-same-funcs} shows two diverse implementations of the same boolean function. If all inputs are logic \texttt{0}, the true output of both circuits is logic \texttt{1}. If a \texttt{stuck-at-1} fault is injected at the same point, the two circuits exhibit different behaviors at the final output \texttt{Y}.

\begin{definition}[Diverse Module]
\label{defn:dm}
Consider any arithmetic logical circuit $m\in M$ modeled as an underlying DAG of vertices (gates) and edges (gate outputs and inputs). 
$M$ is a Diverse Module (DM) iff there exist at least two implementations $m_i, m_j \in M$ such that $m_i \neq m_j$ (differ in at least one DAG edge), and $out(m_i)=out(m_j)$ (deterministic) where $out$ is their output function. 
\end{definition}

\subsubsection{\textbf{Module-level diversity}}
This diversity level aims at building a module artifact as a composition of diverse modules with some dependency between them.
The artifact's output can be dependent on the position/order of faulty or vulnerable module(s). (This is different from module replication in the TMR sense.)
The designer makes use of off-the-shelf modules, diversified at the Gate-level, to create higher order \textit{Composed Module Artifacts (CMAs)} through additional diversity at this layer. 
We formally define a CMA as follows:

\begin{definition}[Composable Module Artifact–CMA]
\label{defn:cla}
Consider an arithmetic logical artifact $L$ with a deterministic specification $S$. An implementation $A$ of $L$ is an ordered set of modules $M_i$ defined as:

\begin{equation}
 \footnotesize
\label{eqn:cla}
    A = (M_1, M_2, ..., M_N) \in \Pi = \{M_i \times M_j \times ... \times M_N\}
\end{equation}

where $\Pi$ is a set of logical circuit modules. $A$ is a Composable Module Artifact (CMA) iff it respects the specification $S$ given a defined connectivity matrix $\Gamma$. 

\begin{equation}
 \footnotesize
\label{array:conn-matrix}
\Gamma_{N \times N} = 
\begin{blockarray}{cccc} 
\begin{block}{[cccc]}
    V_{0,0} & V_{0,1} & \cdots & V_{0,N-1} \\
    V_{1,0} & V_{1,1} & \cdots & V_{1,N-1} \\
    \vdots & \vdots & \ddots & \vdots \\
    V_{N-1,0} & V_{N-1,1} & \cdots & V_{N-1, N-1} \\
\end{block}
\end{blockarray}
\end{equation}

where, 

\begin{equation}
V_{ij} = 
\begin{cases}
  1 & \text{if $a_i$ connected to $a_j$} \\
  0 & \text{otherwise}
\end{cases}
\label{eqn:conn-eqn}
\end{equation}

\end{definition}

An implementation $A$ uses $N$ modules to conform to the specification $S$. The connectivity matrix ($\Gamma$) of size $N \times N$ is a directed acyclic graph (DAG) that defines how Modules are connected with each other. 
For example, to construct an \textit{$N-bit$ Ripple-Carry Adder (RCA)}, the connectivity matrix would simply be the Identity Matrix ($I$) as shown in Eqn.~\ref{eq:adder}.
Similarly, an $N-bit$ multiplier would require $N^2$ AND gates and $2N-1$ adders (either half adders or full adders).


A Composable Module Artifact has modularity properties that can guarantee deterministic logic over diversified modules. We define and prove this in the next Lemma. 

\begin{lemma} (A Diverse CMA is deterministic)
\label{lemma:dv-cma}
A CMA $A_i=(M_i, M_j, ...,M_N)$ whose module implementations $m_k^t$ belong to a diverse module $M_k$ is deterministic.
\end{lemma}
\begin{IEEEproof}
Assume the contrary, that there exist at least two diverse CMA implementations $P=(p_1^a, p_2^b, ..., p_N^z)$ and $Q=(q_1^l, q_2^t, ..., q_N^k)$ with at least one tuple location $x$ having $(p_x^e\neq q_x^e)$, $p_x^e, q_x^e\in M_x$ ($M_x$ being a diverse module), and such that $out(P)\neq out(Q)$ (nondeterministic). Since all module implementations $p_y^f$ and $q_y^g$ belong to a diverse module $M_y$, then it is possible to replace each $p_y^f$ with $q_y^g$ at every index $y$ in CMA $P$. This generates exactly equivalent tuple implementations $P=Q$. Since we assumed $out(P)\neq out(Q)$, then $out(P)\neq out(P)$. This means having one exact implementation of CMA that is not deterministic, which contradicts the CMA definition in Eq.~\ref{eqn:cla}.
Therefore, a diverse CMA is deterministic. \\
\end{IEEEproof}
\

\begin{figure}[t]    
    \centering
        \includegraphics[width=\columnwidth]{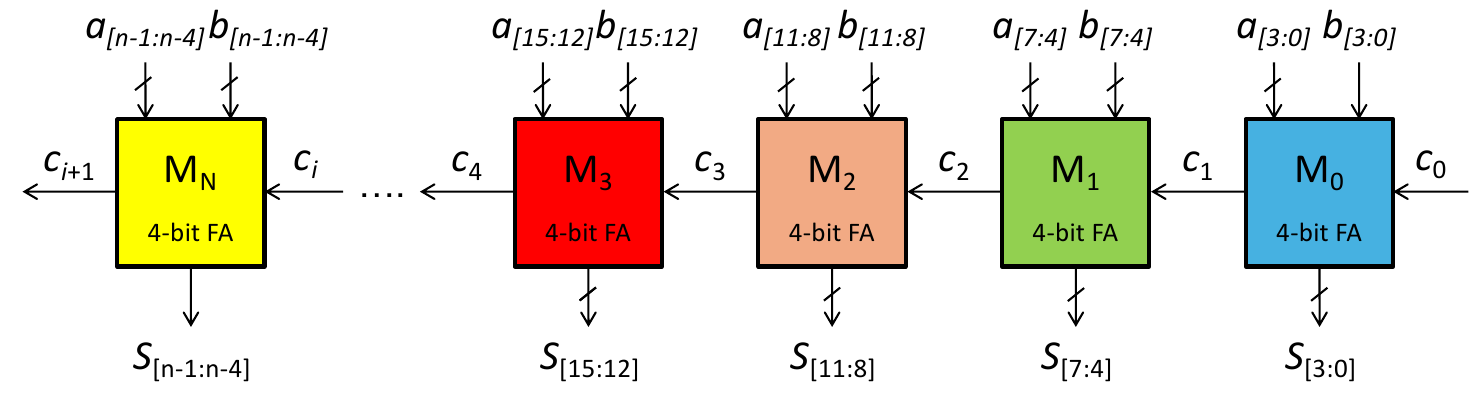}
        \caption{N-bit Ripple-Carry Adder composed of {\it diverse} 4-bit adder modules.}
    \label{fig:add-manifestation}
\end{figure}

Such CMAs can be applied to a wide range of designs such as \textit{N-bit Adders/Subtracters, N-bit Multipliers, N-bit Dividers}, or modern accelerator architectures, e.g. \textit{SHA-256} hashing, Vector units, \textit{Systolic Array} architectures~\cite{jouppi2018motivation}. 
To demonstrate the concept, we discuss two broadly used designs with different complexity: N-bit Adder and \textit{Systolic Array} architecture which is widely used in \textit{Tensor}-based Deep Neural Networks (DNNs).\\

   \begin{equation}
        \label{eq:adder}
         \footnotesize
        Adder\ (\Gamma_{N \times N}) = 
        \begin{blockarray}{ccccc}
         & M_1 & M_2 & \cdots & M_N \\
        \begin{block}{c[cccc]}
            M_0 & {\bf1} & 0 & \cdots & 1 \\
            M_1 & 0 & {\bf1} & \cdots & 0 \\
            \vdots & \vdots & \vdots & \ddots & \vdots \\
            M_{N-1} & 0 & 0 & \cdots & {\bf1}\\  
        \end{block}
        \end{blockarray}
        \end{equation}

{\bf N-bit Ripple-Carry Adder:} Fig.~\ref{fig:add-manifestation} shows an N-bit Ripple-Carry Adder (RCA) composed of smaller modules which can be diverse or homogeneous structures.
The adder is composed of $4-bit$ full adder modules.
The basic resolution of a module can be decided by the designer.
However, smaller modules provide less room for diversity due to the smaller number of gates, while large modules can suffer from worse critical path delays.
A $4-bit$ FA module granularity provides a good trade off for exploring diverse combinations. The corresponding connectivity matrix is an Identity matrix that defines how the modules are connected to realize an arithmetic add/subtract operation. The value of $1$ denotes which corresponding row module is feeding its output to the column module.

\begin{definition}[Homogeneous/Heterogeneous CMA]
An artifact implementation $A_i=(M_i, M_j, ...,M_N)$ is a homogeneous CMA iff $M_x=M_y\ \forall\ M_x,M_y$; otherwise, it is heterogeneous. 
\end{definition}


{\bf 4-bit Array Combinational Multiplier:} The $4-bit$ combinational multiplier in Fig.~\ref{fig:mul-manifestation} is composed of various Half-Adders (HA) and Full-Adders (FAs) which sum partial products.
Composability allows use of diverse implementations of HAs and FAs. For brevity, the $AND$ gates that perform bitwise multiplication are not shown.
In fact, a library of diverse $AND$ gates can also be created to generate partial products through diverse implementations. 
The corresponding connectivity matrix is in Eqn.~\ref{eq:mult}, $ Mul\ (\Gamma_{N \times N})$:

\begin{figure}  
    \centering
        \includegraphics[width=0.9\columnwidth]{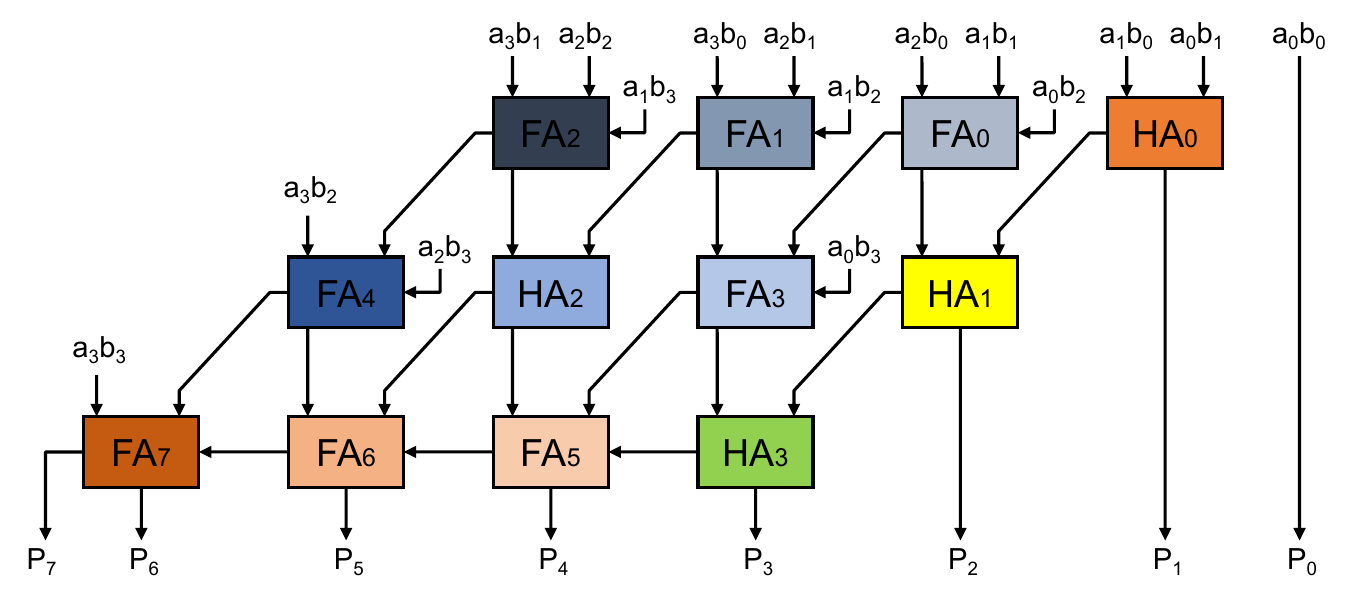}
        \caption{A $4-bit$ Combinational Multiplier composed of {\it diverse} Half-Adders and Full-Adders.}
        \label{subfig:mul}
    \label{fig:mul-manifestation}
\end{figure}

    \begin{equation}
            \label{eq:mult}
         \footnotesize
        \begin{blockarray}{c@{\hspace{1pt}}c@{\hspace{1pt}}c@{\hspace{1pt}}c@{\hspace{1pt}}c@{\hspace{1pt}}c@{\hspace{1pt}}c@{\hspace{1pt}}c@{\hspace{1pt}}c@{\hspace{1pt}}c@{\hspace{1pt}}c@{\hspace{1pt}}c}
        & HA_1 & HA_2 & HA_3 & FA_1 & FA_2 & FA_3 & FA_4 & FA_5 & FA_6 & FA_7\\
        \begin{block}{c[ccccccccccc]}
            HA_0 & {\bf1} & 0 & 0 & 0 & 0 & 0 & 0 & 0 & 0 & 0 \\
            HA_1 & 0 & 0 & {\bf1} & 0 & 0 & 0 & 0 & 0 & 0 & 0 \\
            HA_2 & 0 & 0 & 0 & 0 & 0 & 0 & 0 & {\bf1} & {\bf1} & 0 \\
            HA_3 & 0 & 0 & 0 & 0 & 0 & 0 & 0 & {\bf1} & 0 & 0 \\
            FA_0 & {\bf1} & 0 & 0 & 0 & 0 & {\bf1} & 0 & 0 & 0 & 0 \\
            FA_1 & 0 & {\bf1} & 0 & 0 & 0 & {\bf1} & 0 & 0 & 0 & 0 \\
            FA_2 & 0 & {\bf1} & 0 & 0 & 0 & 0 & {\bf1} & 0 & 0 & 0 \\
            FA_3 & 0 & 0 & {\bf1} & 0 & 0 & 0 & 0 & {\bf1} & 0 & 0 \\
            FA_4 & 0 & 0 & 0 & 0 & 0 & 0 & 0 & 0 & {\bf1} & {\bf1} \\
            FA_5 & 0 & 0 & 0 & 0 & 0 & 0 & 0 & 0 & {\bf1} & 0 \\
            FA_6 & 0 & 0 & 0 & 0 & 0 & 0 & 0 & 0 & 0 & {\bf1} \\
        \end{block}
        \end{blockarray}
    \end{equation}

The matrix highlights how the modules are connected  to realize a multiplication operation.
It can also be observed that most of the elements in the matrix are zero i.e., sparse. We foresee it to be a typical structure of a connectivity matrix.

\begin{lemma}
\label{lemma:heterogeneous-cma}
There exists an N-bit Multiplier implementation that is a heterogeneous CMA.  Proof in Appendix~\ref{lemma:heterogeneous-cma}.
\end{lemma}

\begin{figure}[t]    
    \centering
        \includegraphics[width=1\linewidth]{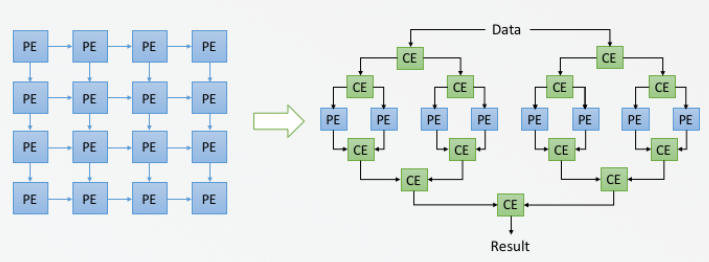}
        \caption{Composable mesh and hierarchical structures of Systolic Arrays~\cite{lu2017ai}, composed of (diverse) Processing Elements (PE) and Collection Elements (CE).}
    \label{fig:systolic-arrays}
\end{figure}

{\bf Systolic Arrays:} Systolic arrays\cite{kung1979systolic, quinton1983systematic} are a class of parallel computing architectures designed to efficiently execute matrix computations, which are fundamental to many scientific and engineering applications~\cite{xu2023survey, snopce2020mapping} including signal processing~\cite{snopce2020mapping}, image processing, and machine learning~\cite{xu2023survey}. 
Based on the Systolic architecture, Google proposed Tensor Processing Units (TPU)~\cite{GoogleTPU2017, jouppi2018motivation}, a custom ASIC to accelerate DNN applications. 
A typical Systolic array design consists of a 2-Dimensional grid of Processing Elements (PE), a controller and the on-chip memory/buffer.
Fig.~\ref{fig:systolic-arrays} depicts two examples of composable Systolic Array architectures using Processing Elements (PEs) and Collection Elements (CE)~\cite{lu2017ai}.
Leveraging the design of diverse adders and multipliers, diverse PEs can be designed to enhance the resilience of systolic arrays.\\

{\bf Further Use Cases and Applications:} Similar to these Adder, Multiplier and Systolic arrays CMA implementations, there is a large variety of other useful arithmetic designs that are composable in this manner~\cite{shao2015aladdin}.
The artifacts we target in this work are basic logic circuits that are widely used, making this approach applicable in many use cases.


Nevertheless, given our threat model, our work may be even more applicable in ASIC or FPGA designs, where the area dedicated to arithmetic operations is significantly higher.
More promising use cases may target accelerators for ML algorithms or cryptographic functions, e.g., \textit{SHA256} or Cryptocurrency miners~\cite{calvao2019crypto}.

Modern cryptographic and Machine Learning (ML) algorithms rely heavily on parallelized datapaths with ample arithmetic modules. \textit{SHA-256} is an integral part of cryptographic algorithms to achieve high level of security~\cite{padhi2017optimized}. It consists of several rounds of rotate and XOR operations, making it a rich target for applying the composability principle by choosing diverse implementation for each operation.

\subsubsection{\textbf{Artifact-level diversity}}
This third level of diversity makes use of multiple \textit{replicas} of diverse CMAs running simultaneously, the outputs of which are computed using a majority voter. This is similar to TMR with an important difference: common mode failures are reduced significantly, since diverse CMAs of diverse modules can be used by composition. Visual examples are shown in Fig.~\ref{fig:replicas-manifestation}. Our evaluation in Section~\ref{sec:results-discussions} shows that this level is key to thwart Distribution and Compound attacks in particular.

Furthermore, due to the coarse granularity, the placement of these replicas can be directed to leverage some resilience to Zonal attacks. Combining placement with diversity at different levels boosts the resilience of the system against all considered attacks: Distribution, Zonal, and Compound, as we convey in the evaluations in Section~\ref{sec:results-discussions}.

%% file: resilogic-framework.tex
\section{The \acronym Framework}
\label{sec:resilogic-framework}
This section presents the \acronym framework that is used by the designer to implement resilient designs based on the diversity concepts introduced in Section~\ref{sec:diversity-for-resilience}. The framework defines a mechanism to construct circuits while: (1) maintaining deterministic behavior across the entire design despite replication and diversity, and (2) ensuring resiliency to the attacks outlined in Section~\ref{sec:attack-model}.

In \acronym, the ultimate goal is to produce a resilient circuit corresponding to a CMA 
following Definition~\ref{defn:cla}. The process follows three main phases summarized in Algorithm~\ref{alg:proposed-framework}, with the details discussed next.

\subsection{Building Diverse Modules via Gate-diversity} 
This phase aims at identifying the different candidate module implementations $m_i$ for use in a CMA implementation $A = (m_1, m_2, ..., m_N)$; e.g., $m_i:=$4-bit-adder in our example. The aim is to generate diverse modules of $m_x^l,\ m_x^t \in M_x$, such that  $out(m_x^t)=out(m_x^z)$, where $out$ is the output function, but the structure differs in at least one logic gate, i.e., $m_x^t\neq m_x^z$, thus generalizing the diverse CLA to $A_i=(M_i, M_j, ...,M_N)$ in the next phase. 
Therefore, the designer enumerates off-the-shelf and/or custom implementations of these candidate modules with the criteria defined in the Diverse Module Definition~\ref{defn:dm}: all module implementations of $M_x$ are deterministic, i.e., Determinism is, hence, guaranteed by definition.
In the case of a homogeneous CMA, a single module type is used across the entire tuple.
Heterogeneous CMAs will however be built from diverse modules. To automate this process, we utilize a recent technique for diversifying logic based on E-Graphs~\cite{2021-egg}.

\begin{figure}[ht]   
    \centering
    \begin{subfigure}{0.19\textwidth}    
        \includegraphics[width=.9\textwidth]{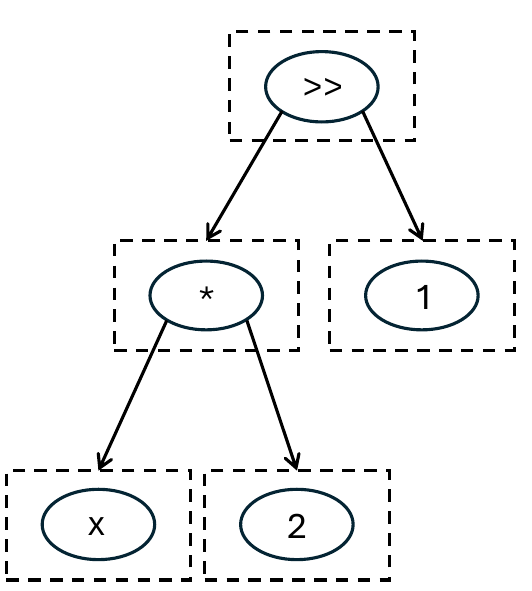}
        \caption{Initial E-graph: \\ $(x \times 2) >> 1$}
        \label{subfig:egraph-figa}
   \end{subfigure}%
    \begin{subfigure}{0.19\textwidth}
       \includegraphics[width=1\textwidth]{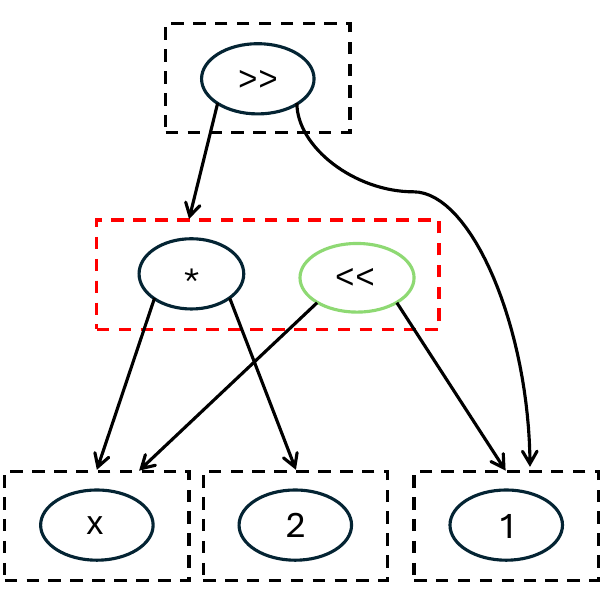}
        \caption{Apply: \\ $(x \times 2) \rightarrow x << 1$}
        \label{subfig:egraph-figb}
   \end{subfigure}%
\caption{E-graph rewriting for standard integer arithmetic. Dashed boxes
represent e-classes of equivalent expressions. Green nodes represent newly
added nodes. Red dashed boxes highlight which e-class has been modified. The arcs are between e-nodes and e-classes.}
\label{fig:egraph-example}
\end{figure}

\textit{Example.} Fig.~\ref{fig:egraph-example} highlights the application of the E-Graph in Fig.~\ref{subfig:egraph-figa} through a \texttt{rewrite} with equivalence operation $(2 \times x)$ is equivalent to $(x << 1)$ to generate an equivalent graph in Fig.~\ref{subfig:egraph-figb}, with equivalent operations grouped in the same e-class (red dotted box). Consequently, two mathematically equivalent relations can then be extracted (black dotted box): 1)  $(x \times 2) >> 1$ and 2) $(x << 1) >> 1$ with different resiliency profiles.

\subsubsection{E-graph transformations}
An E-Graph or equivalence graph
is a data structure that compactly represents equivalence relations in the form of \textit{Abstract Syntax Trees} (AST)~\cite{2021-egg}. 
E-Graphs have been mainly used in the optimization of large datapath and arithmetic circuits~\cite{coward2022automatic, coward2024rover, coward2024combining, ustun2022impress} and technology aware synthesis~\cite{chen2024esyn}. In this work, we use them to generate diverse circuit designs using equivalence \textit{Transformations}.
A \texttt{rewrite} is a simple AST representing an equivalence operation that is successively applied on an E-Graph to generate an equivalent graph compactly with in the same graph.
\texttt{Extraction} refers to selecting AST(s) from E-Graph based on a certain cost function. 
The default cost functions are \texttt{ASTSize} and \texttt{ASTDepth}, E-Graph tool constructs an optimal AST as a result of minimizing these cost functions. 

\subsubsection{Diversity transformations}
We automate the generation of diverse logic circuits using new specifications for \textit{rewrite} and \textit{extraction}. We proposed in Table~\ref{table:resilogic-rewriting} a \texttt{rewrite} list definition rules for different classes of operations. These rules are used attractively to generate an enormous number of equivalence versions. Afterwards, extraction is defined such that versions have at least one structural difference in its implementation. The \textit{rewrite} and \textit{extraction} transformations ensured versions that are functionally equivalent but structurally different.

\begin{table}[t]
\setlength{\tabcolsep}{1pt}
\caption{\acronym Rewriting Rules}
\label{table:resilogic-rewriting}
\begin{tabular}{ll}
\hline
\textbf{Class}                           & \textbf{Boolean Rewriting Rules}                                       \\ \hline
\multirow{6}{*}{\textbf{Complements}}    & $a * 0 \implies (a * \overline{a})$                                    \\ 
                                         & $a*1 \implies \overline a*{(\overline{a}*a)+(\overline{a}*a)}$         \\
                                         & $a + 1 \implies (a + \overline{a})$                                    \\
                                         & $a+0 \implies \overline a+{(\overline{a}*a)+(\overline{a}*a)}$         \\
                                         & $a*\overline{a} \implies (\overline{a}*a)+(\overline{a}*a)$            \\
                                         & $a+\overline{a} \implies \overline{(\overline{a}*a)+(\overline{a}*a)}$ \\ \hline
\multirow{2}{*}{\textbf{Covering}}       & $a*(a+b)\implies (a*a) + (a*b)$                                        \\
                                         & $a+(a*b)\implies (a*b) + (a*\overline{b})$                             \\ \hline
\multirow{2}{*}{\textbf{Combining}}      & $(a*b)+(a*\overline{b}) \implies (a*b)+(a*\overline{b})$               \\
                                         & $(a+b)*(a+\overline{b}) \implies (a+b)*(a+\overline{b})$               \\ \hline
\multirow{2}{*}{\textbf{Idempotency}}    & $a*a \implies (a*a)+(a*\overline{a}$                                   \\
                                         & $a+a \implies (a*a)+(a*\overline{a}$                                   \\ \hline
\multirow{2}{*}{\textbf{Commutativity}}  & $a*b \implies b*a$                                                     \\
                                         & $a+b \implies b+a$                                                     \\ \hline
\multirow{2}{*}{\textbf{Associativity}}  & $(a*b)*c \implies (a*(b*c))$                                           \\
                                         & $(a+b)+c \implies (a+(b+c))$                                           \\ \hline
\multirow{2}{*}{\textbf{Distributivity}} & $a*(b+c) \implies (a*b)+(a*c)$                                         \\
                                         & $a+(b*c) \implies (a+b)*(a+c)$                                         \\ \hline
\multirow{2}{*}{\textbf{Consensus}}      & $(a*b)+(\overline{a}*c)\implies (a*b)+(\overline{a}*c)+(b*c)$          \\
                                         & $(a+b)*(\overline{a}+c)\implies (a+b)*(\overline{a}+c)*(b+c)$          \\ \hline
\multirow{2}{*}{\textbf{De-Morgan}}      & $\overline{(a*b)}\implies \overline{a}+\overline{b}$                   \\
                                         & $\overline{(a+b)}\implies \overline{a}*\overline{b}$                   \\ \hline
\end{tabular}
\end{table}

\begin{figure}
    \centering
    \includegraphics[width=\linewidth]{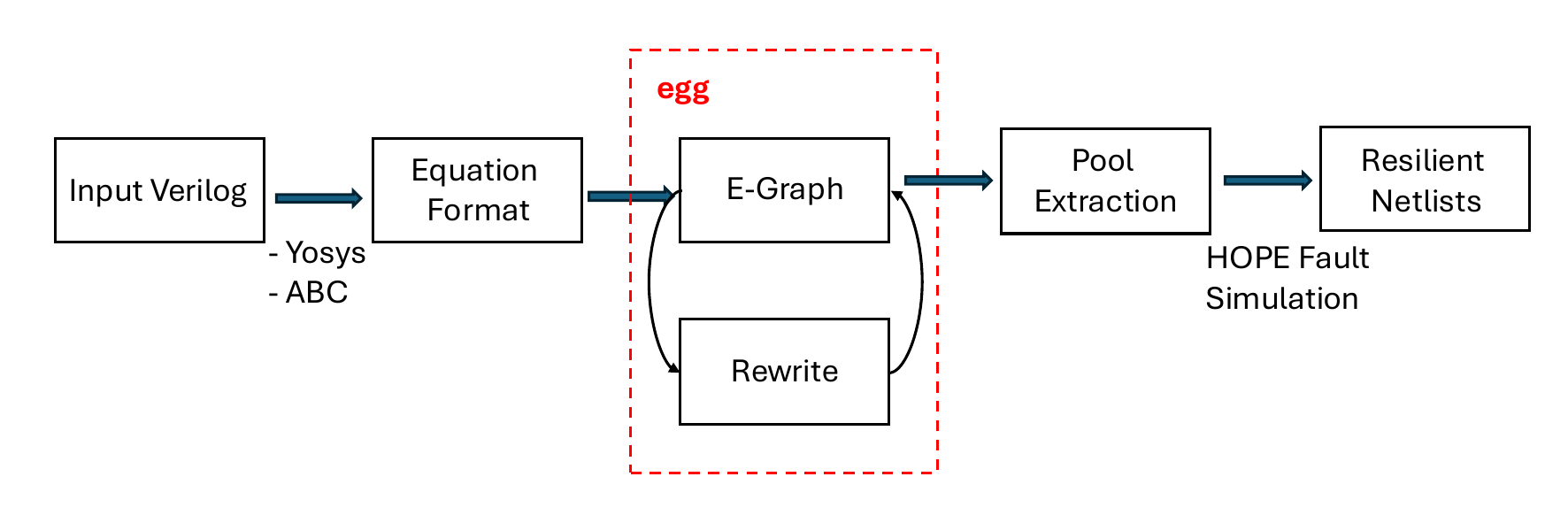}
    \caption{E-graph flow diagram.}
    \label{fig:egraph-framework}
\end{figure}

\subsubsection{E-graph diversity implementation}
Practically, the implementation of the above process is depicted in the E-Graph flow in Fig.~\ref{fig:egraph-framework} An input circuit represented in the \textit{Verilog} format is synthesized and converted into equation format which represents the circuit in the \textit{AND}, \textit{OR} and \textit{NOT} format. Next, the \texttt{rewrite} operations are applied until saturation, and finally a pool of diverse implementations is extracted. Finally, a fault simulation tool, like \textit{HOPE}~\cite{Hope1996}, is utilized to quantify the fault coverage of pooled \textit{netlists} and perform further pruning to select the candidates that demonstrate low fault coverage i.e., high resiliency.

Transformations in E-Graph consist of two main steps 1) \texttt{rewrite} and 2) \texttt{extraction}. 
We use Yosys/ABC logic synthesis~\cite{wolf2013yosys, brayton2010abc} flow to synthesize a benchmark circuit into an equivalent \texttt{equation} format that could be fed to the E-Graph tool for design space exploration. 

\begin{table}
\centering
\caption{E-Graph Diversity Generation Benchmarks.}
\label{table:egg-benchmarks}
\begin{tabular}{|l|c|c|c|}
\hline
\multicolumn{1}{|c|}{\textbf{Circuit}} & \textbf{I/O} & \textbf{\begin{tabular}[c]{@{}c@{}}Area\\ (Gates)\end{tabular}} & \textbf{\begin{tabular}[c]{@{}c@{}}Prob. Failure \\ ($P_f$)\end{tabular}} \\ \hline
MAC                                    & 16/8         & 121                                                                   & 0.99                                                                 \\ \hline
3\_3 (genmul)                          & 6/6          & 46                                                                    & 0.96                                                                 \\ \hline
5\_5 (genmul)                          & 10/10        & 167                                                                   & 0.99                                                                 \\ \hline
b12 (ITC99)                            & 126/127      & 1069                                                                  & 0.87                                                                 \\ \hline
c432 (LGSynth89)                       & 36/7         & 247                                                                   & 0.95      
\\ \hline
c2670 (ISCAS85)                        & 233/139      & 837                                                                   & 0.95                                                                 \\ \hline
\end{tabular}
\end{table}

\begin{figure}[t]
    \centering
        \begin{subfigure}[b]{0.45\textwidth}
    \includegraphics[width=\linewidth]{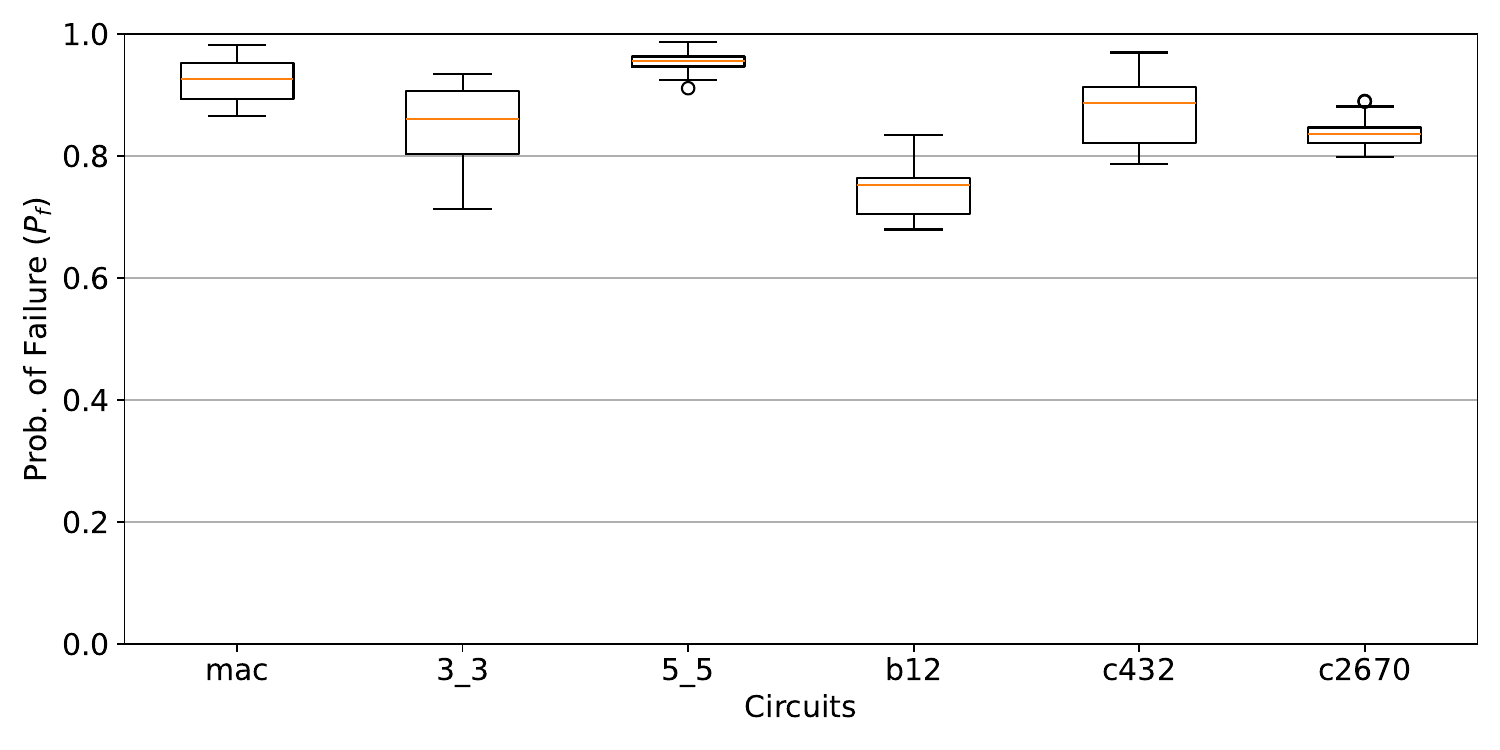}
    \caption{Resilience}
    \label{fig:egg-resilience}
\end{subfigure}%

    \begin{subfigure}[b]{0.45\textwidth}
    \centering
    \includegraphics[width=1\linewidth]{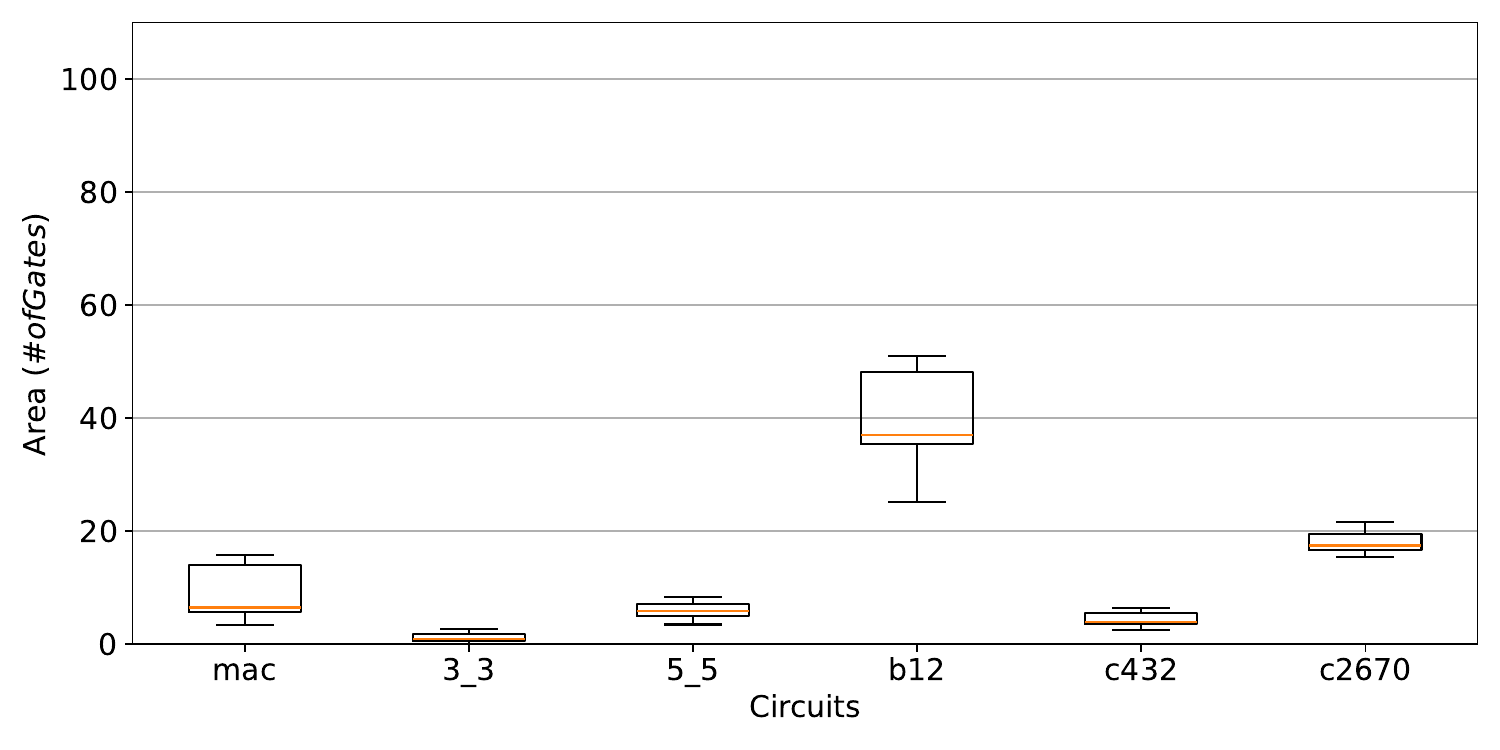}
    \caption{Area}
    \label{fig:egg-area}
    \end{subfigure}%
\caption{Resilience and area ranges of diverse generated artifacts of benchmarks with Min-Max normalization.}
\end{figure}

\subsubsection{E-graph diversity evaluation}
\label{subsec:resilience-with-egg}
The effectiveness of E-Graph based approach to generate diverse and resilient circuits is evaluated on the benchmark circuits shown in Table~\ref{table:egg-benchmarks}. 
The benchmarks are a mix of circuits from LGSynth~\cite{lisanke1988logic, yang1991logic}, ITC~\cite{corno2000rt}, and ISCAS85~\cite{brglez1989combinational} benchmark suites. 
Additionally, an open source divider from OpenCores is also considered for evaluation. 
The considered benchmarks are: \texttt{genmul}~\cite{mahzoon2021genmul} that generates two multipliers (3x3 and 5x5); \texttt{mac} benchmark that corresponds to the Multiply-Accumulate operation unit of a typical \textit{Systolic} array; LGSynth~\cite{lisanke1988logic, yang1991logic}, ITC~\cite{corno2000rt}, and ISCAS85~\cite{brglez1989combinational} benchmark suites. 
The \texttt{rewrite} (Table~\ref{table:resilogic-rewriting}) and \texttt{extract} methods of E-Graphs are implemented in \texttt{Rust} to interface with the \texttt{egg} library.

Fig.~\ref{fig:egg-resilience} and~\ref{fig:egg-area} highlight the resilience and area overheads of the above benchmarks. For benchmarks \texttt{3\_3} and \texttt{c432}, the \texttt{egg} tool was able to generate diverse replicas with varying resilience ($P_f$) levels at low area overhead. The area overhead increased in the cases of \texttt{mac}, \texttt{c2670} and \texttt{b12}.

\subsubsection{Observations and analysis}
We observed that the resilience and corresponding area profiles of the egg tool are sensitive to the \texttt{rewrite} rules in Table~\ref{table:resilogic-rewriting}. New rules added or existing ones modified can lead to different implementations and area/resilience profiles. However, since optimizing this process is orthogonal to this work, the obtained diversity was satisfactory to (1) demonstrate our proof of concept and (2) generate enough diversity for the rest of the evaluation. In addition, one can notice that the probability of failure is not significantly improved with this level of fine-grained diversity. Nevertheless, combined with the other levels of diversity presented in the next sections, we show that the probability of failures is reduced significantly. An interesting research would be to optimize the E-Graph transformations to generate more diverse circuits at a lower area cost.

\subsubsection{Synthesis Impact on Resilience}

\begin{figure*}[ht]
    \begin{subfigure}{0.33\textwidth}
            \includegraphics[width=\textwidth, height=0.2\textheight]{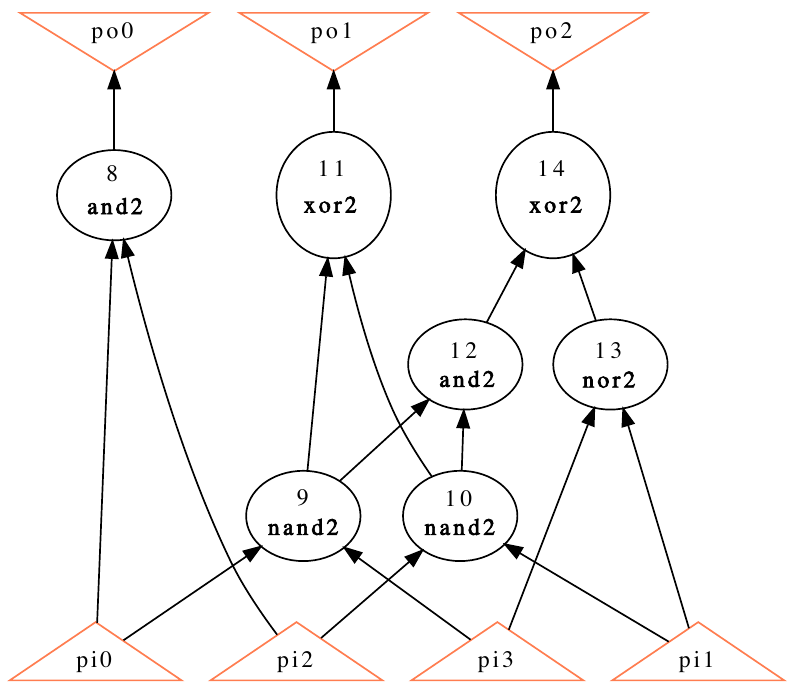}
            \caption{Example circuit (7 Logic Gates)\\$P_f=1$}
            \label{fig:appendix-sample-circuit}
    \end{subfigure} 
    \begin{subfigure}{0.33\textwidth}            
            \includegraphics[width=\textwidth, height=0.2\textheight]{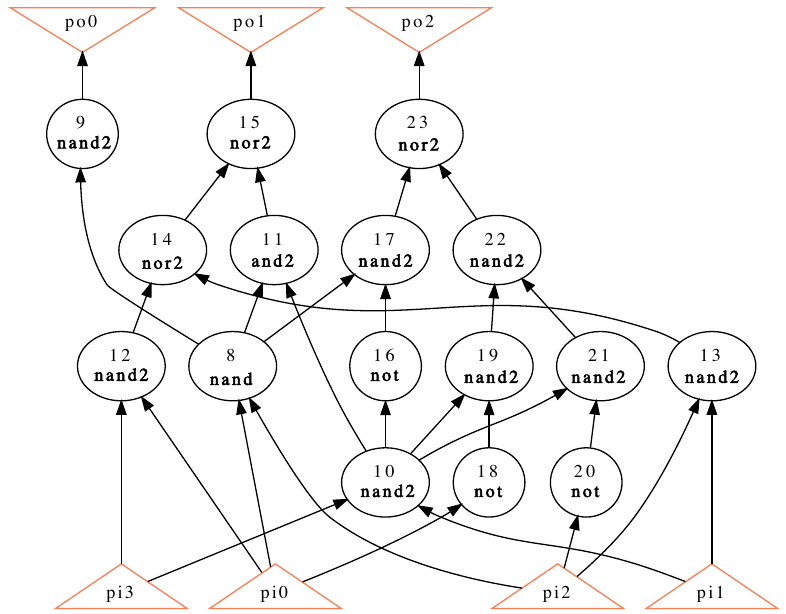}
            \caption{With Optimization (16 Logic Gates)\\$P_f=0.84$}
            \label{fig:appendix-sample-wopt}
    \end{subfigure}        
    \begin{subfigure}{0.33\textwidth}
           \includegraphics[width=\textwidth, height=0.2\textheight]{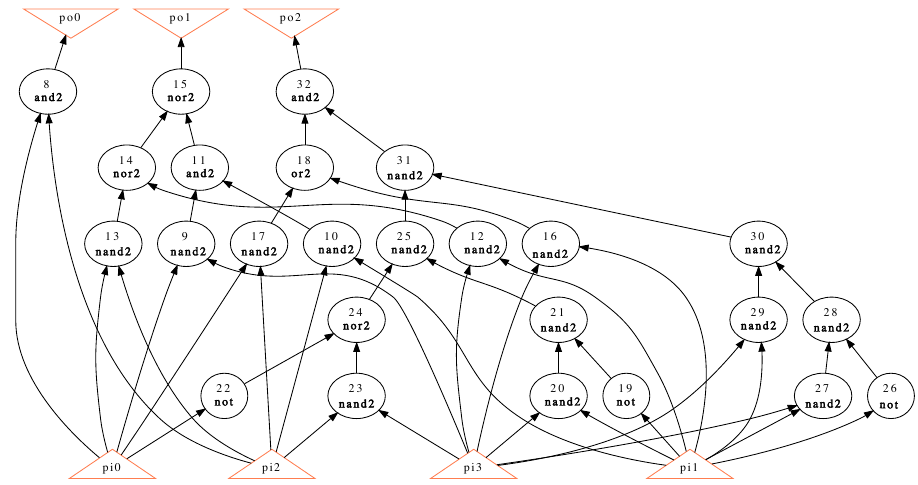}
            \caption{Without Optimization (25 Logic Gates)\\$P_f=0.74$}
            \label{fig:appendix-sample-woopt}
    \end{subfigure}%
    \caption{Synthesis impact on  sample circuit for diversity. $P_f$ is computed using HOPE~\cite{Hope1996} parallel fault simulation tool.}
        \label{fig:appendix-synth-impact-example}
\end{figure*}

The inherent nature of Electronic Design Automation (EDA) synthesis tools is to optimize circuits to meet the area, power and delay constraints. To achieve resilience, additional gates have to be added that doesn't change the function of circuit but increase the area and delay. Since E-Graph works on the ASTs of a given language, therefore, it relies on parser and lexer for text processing. Any redundancy in the resulting E-Graphs of diverse replicas will 1) be removed by the parser, or 2) removed by the synthesis tools.To avoid these optimizations, following steps can be followed:  1) rename each redundant gate with a different tag, 2) declare all redundant gates as primary inputs to the circuit, 3) synthesis will not perform any optimizations, and 4) remove redundant gates from the primary inputs list. Following these steps, it is observed that high resilience can be achieved, but at the cost of exorbitant area. 

Fig.~\ref{fig:appendix-synth-impact-example} illustrates the impact of synthesis on digital circuits. Example circuit (Fig.~\ref{fig:appendix-sample-circuit}) consists of 7 logical gates with $P_f=1$. Applying E-Graph rewrite and extract operations with optimizations i.e., not tagging each gate, the resulting area after synthesis is 16 gates with the corresponding $P_f=0.84$, as shown in Fig.~\ref{fig:appendix-sample-wopt}. However, if each gate is tagged i.e., no optimization is applied, synthesis of the circuit will result in 25 logical gates with $P_f=0.74$, as shown in Fig.~\ref{fig:appendix-sample-woopt}. This simple example highlights the area is increased by more than 3x if no optimization is applied. For larger circuits, we observed that the resulting area becomes exorbitantly high (with corresponding significant reduction in $P_f$) making the final design unfeasible for practical applications. It will be interesting to investigate the impact of E-Graphs to generate diverse replicas at different levels of digital circuits abstraction.

\subsection{Intra-diversity: building a CMA out of Diverse Modules} 
This phase leverages the generated diverse modules $M_i$ to build a resilient diverse CMA $A_i=(M_i, M_j, ...,M_N)$. The designer starts with the standard implementation of the CMA using any generated module implementations in the previous phase.
A conservative option is to use the most resilient CMA implementation $A = (m_1, m_2, ..., m_N)$.
In general, this depends on the policy considering the resilience, cost, space, and time tradeoffs.
To create diverse CMAs, the designer iterates on the tuple to replace a module implementation $m_i^t$ with another generated one $r_i^t\in M_i$.
Diverse CMAs can be generated either by changing a single module implementation at index $i$ or multiple modules in CMA $A_i$.

\begin{algorithm}[t]
\footnotesize
\caption{\acronym Diversification}
\label{alg:proposed-framework}
\begin{algorithmic}[1]
    \Require {\bf$\Pi$}: Set of diverse modules $M_i$
    \Require {\bf$\Gamma$}: Connectivity Matrix
    \State {\bf$Res$} : Required resilience range of modules
    \State {\bf$I$} : Intra-diversity factor
    \State {\bf$D$} : Inter-diversity factor
    \State {\bf$R$} : Number of fault resilient replicas
    \State {\bf$N$} : Required bit-width of artifact    
    \State $K\gets FALSE$
        \While{($K \ne TRUE$)}   
            \For{$(i = 1 \to R)$} \Comment{Create $R$ replicas}
                \State Select modules satisfying $N$ based on $I$ \& $Res$ factor
            \EndFor
            \If{($D$ condition satisfied among replicas)}
                \State Apply $\Gamma$ to realize replicas operations
                \State Connect replicas to the majority voter
                \State $K\gets TRUE$                          
            \EndIf            
        \EndWhile
\end{algorithmic}
\end{algorithm}



\subsection{Inter-diversity: building a replicated artifact} 
This phase uses TMR-style replication of the \textit{diverse} CMAs.
The designer starts by selecting one CMA implementation and then replicates it into three replicas whose outputs are connected to a majority voter.
(One can scale the replication to any numbers.)
The designer then iterates over the three tuples to replace the CMA implementations with diverse homogeneous or heterogeneous ones.
The latter is preferable as this utilizes Inter-diversity between CMAs to defend against Distribution and compound attacks by reducing common mode failures.
To defend against Zonal attacks, we leverage the coarse-grained nature of CMA replication to define a resilient-aware fixed placement of redundant modules in the chip design.



\subsection{\acronym Configurations}
The above multi-level diversity is expressed through configurations that combine the Intra- and Inter-diversity.
We define these configurations as follows.
We denote the Intra-diversity of a CMA by $\alpha$.
This refers to the number of different modules used in the CMA composition.
Since different CMAs in a replicated artifact can have different Intra-diversity $\alpha$, we model these as a tuple $\mathbf{I}$ whose dimension $|\mathbf{I}|$ is the replication factor. In this case, each index $I_i$ refers to the Intra-diversity value for the implementation $A_i$ of replica $i$.
The $\beta_\mathbf{I}$ factor for a given tuple $\mathbf{I}$, describes the Inter-diversity among the replica CMAs i.e., how diverse they are with respect to each other.
$\beta_\mathbf{I}$ is computed by taking the symmetric difference among replicated CMAs, as shown in Eqn.~\ref{eqn:d}.
To compute $\beta_\mathbf{I}$ for $|\mathbf{I}|>2$, Eqn.~\ref{eqn:d} can be applied iteratively.

\begin{equation}
\label{eqn:d}
    \beta_\mathbf{I} = (A_i - A_{i+1}) \cup (A_{i+1} - A_{i})
\end{equation}

\begin{figure}[t]    
    \begin{subfigure}{0.19\textwidth},
    \centering
        \includegraphics[width=.8\textwidth]{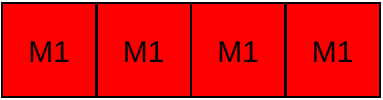}
        \caption{$\mathbf{I}=(1), \beta_\mathbf{I}=()$}
        \label{subfig:i1}
   \end{subfigure}%
   \rulesep
    \begin{subfigure}{0.19\textwidth}
     \centering
        \includegraphics[width=.8\textwidth]{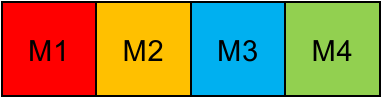}
        \caption{$\mathbf{I}=(4), \beta_\mathbf{I}=()$}
        \label{subfig:i4}
   \end{subfigure}%
   \rulesep
    \begin{subfigure}{0.19\textwidth}
     \centering
        \includegraphics[width=.8\textwidth]{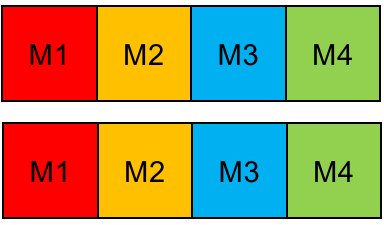}
        \caption{$\mathbf{I}=(4,4),\beta_\mathbf{I}=0$}
        \label{subfig:i4d0}
    \end{subfigure}     
    \rulesep
    \begin{subfigure}{0.19\textwidth}
     \centering
        \includegraphics[width=.8\textwidth]{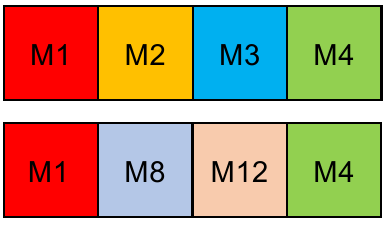}
        \caption{$\mathbf{I}=(4,4),\beta_\mathbf{I}=4$}
        \label{subfig:i4d4}
    \end{subfigure}%
    \rulesep
    \begin{subfigure}{0.19\textwidth}
     \centering
        \includegraphics[width=.8\textwidth]{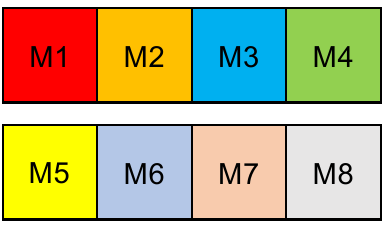}
        \caption{$\mathbf{I}=(4,4),\beta_\mathbf{I}=8$}
        \label{subfig:i4d8}
    \end{subfigure}
    \caption{Replicas manifestation with varying $\alpha\ \&\ \beta$ values.}
    \label{fig:replicas-manifestation}
\end{figure}

Fig.~\ref{fig:replicas-manifestation} shows few example configurations for different values of $\mathbf{I}$ and $\beta_\mathbf{I}$. (The terms CMAs and replicas will be used interchangeably.) In Fig.~\ref{subfig:i1}, a single replica is created with a single module $M_1$ i.e., $\mathbf{I} = (1)$. Fig.~\ref{subfig:i4} shows a single replica consisting of a set of four different modules $\{M1,M2,M3,M4\}$ i.e., $\mathbf{I}=(4)$ and $|\mathbf{I}|=1$. Fig.~\ref{subfig:i4d0} shows a case where $|\mathbf{I}|=2$ with $\mathbf{I}=(4,4)$ i.e., two replicas are created with four diverse modules, however, they are the same as each other, hence $\beta_\mathbf{I}=0$. Fig.~\ref{subfig:i4d4} demonstrates a case where four distinct modules are used to create two replicas ($\mathbf{I}=(4,4)$) with two common modules $M1$ and $M4$, the inter-diversity ($\beta_\mathbf{I}$), computed using Eqn.~\ref{eqn:d}, is $4$.
Lastly, Fig.~\ref{subfig:i4d8} consists of two replicas created with configuration $\mathbf{I}=(4,4), \beta_\mathbf{I}=8$, which implies each replica is composed of four different modules and there are no common modules between them, therefore, they are completely diverse.

\subsection{\acronym Scalability}
\label{subsec:scalability}
Let's say that the basic unit of computation is defined as a module $M$ and the collection of such modules be denoted by $M_i$. 
A replica $R_i$ is a union of modules $M_i$ arranged and connected together in a certain way. The number of modules in a replica is proportional to the size and functionality of an artifact aimed to be realized. A module $M_i$ is a basic unit of function and a set of such modules $\Pi$ represents diverse implementations of some function or a set of modules with multiple functions. A replica $R_{x}$ with required bit width $x$ and $|M_i|$ being the bit width of the module $M_i$, can be defined as:

\begin{equation}
\label{eqn:replica-def}
    R_{x} = \bigcup_{j=1}^{x/|M_i|} M_i \qquad \forall M_i \in \Pi = [M_1, ... ,M_N]
\end{equation}

Large SoC designs typically comprise interconnected components with varying functions and levels of design complexity. \acronym focuses on operating at the component-level abstraction. This approach significantly reduces the complexity of designing large SoCs.
However, the time complexity of generating a diverse system with $N$ replicas that meet the $I$ and $D$ factors in Algorithm~\ref{alg:proposed-framework} is directly proportional to these factors.

Design of diverse chips is not a panacea but an attempt to increase the confidence in system resilience against the CMF. With diverse modules at our disposal, design of resilient circuits would then be an amalgamation of these modules along with the connectivity matrix $\Gamma$ to realize a function. Through the proposed approach, the set of diverse modules can be created leading to the creation of larger diverse artifacts. For fault resilience, mere diversity in a single replica is not enough, a replicated system with Triple Modular Redundancy (TMR) has to be created to: 1) tolerate single fault, 2) detect CMF. TMR ensures to tolerate a single fault, however, a common vulnerability can be exploited to bring the whole system down to its knees. That said, Algorithm~\ref{alg:proposed-framework} describes the steps to create resilient systems in the wake of persistent and dynamic threat. The algorithm creates fault resilient replicas by satisfying the intra and inter diversity conditions.

%% file: results.tex
\section{Results \& Discussions}
\label{sec:results-discussions}

\subsection{Experimental Settings}
Our evaluation aims at measuring the resilience of \acronym's generated diverse circuits to the Distribution, Zonal, and Compound attacks. Since resilient circuits entail redundancy, we also measure the overhead on time, space, and power. 

\subsubsection{Measuring common mode failures (CMF)}
Quantifying diversity between two digital circuits is typically done through fault injection at the gate level and measuring the probability of failures. Mitra et. al.~\cite{mitra2004efficient} used random fault injections which does not fully capture the failure space. To address this, we have built a 
testbed CMF engine that systematically injects faults into gate locations of two candidate diverse circuits following a fault campaign as discussed in Section~\ref{sec:attack-model}. 
The output is then compared to a reference golden circuit where faults are injected. Eqn.~\ref{eqn:cmf-det} shows that for the CMF output to be true, all outputs $O_k$ from the replicas must be equal to each other and different from the reference output $RO$.

\begin{equation}
\label{eqn:cmf-det}
    CMF \equiv O_i=O_j \quad AND \quad O_1 \neq RO \quad \forall\ i \neq j
\end{equation}

\begin{figure}[tb]
    \centering
    \includegraphics[width=0.8\linewidth]{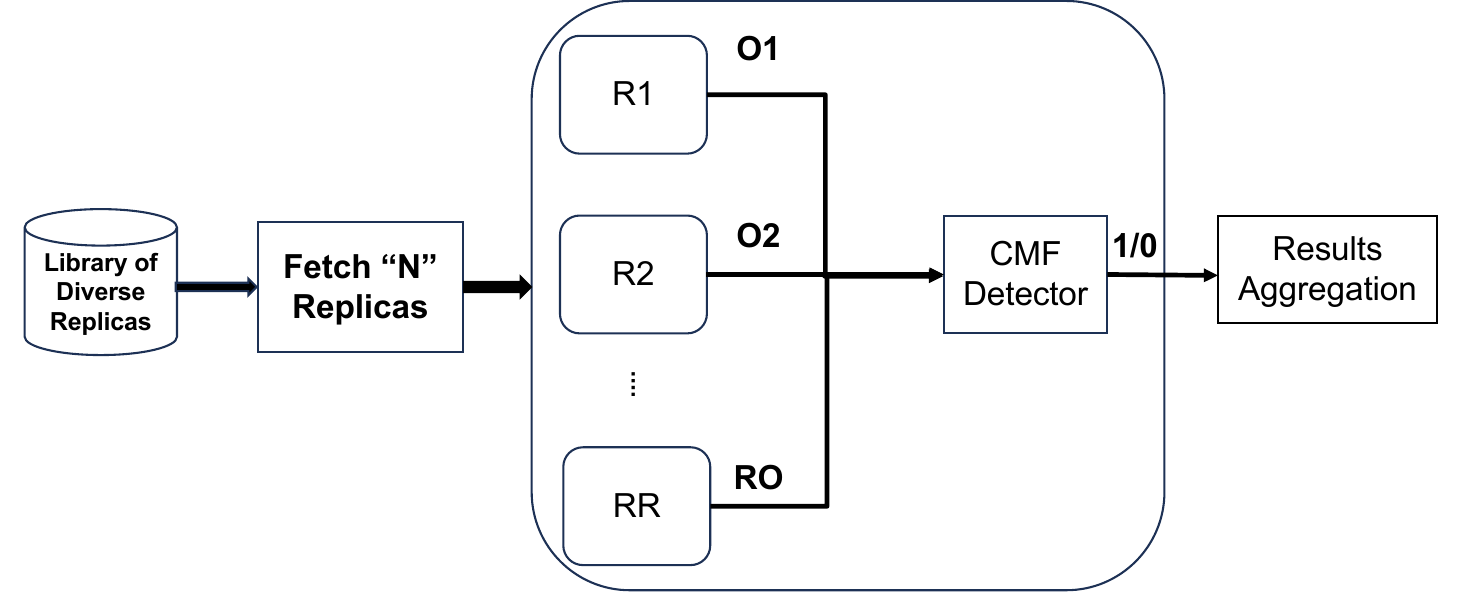}
    \caption{CMF Computing Engine Architecture.}
    \label{fig:cmf-compute-engine}
\end{figure}

Fig.~\ref{fig:cmf-compute-engine} shows the block diagram of the proposed common mode failure computing engine. It fetches a set of diverse replicas from the library generated by the method discussed in Section~\ref{sec:resilogic-framework}.
No fault is injected in $RR$, which serves as a reference circuit producing the golden output. 
The number and sites of faults to be injected depends on the selected fault campaign as discussed in Section~\ref{sec:attack-model}.
Eqn.~\ref{eqn:cmf-det} shows that for the CMF output to be true, all outputs from the replicas must be equal to each other and different from the reference output $RO$. The steps to quantify the CMF are inspired from~\cite{Sheikh2016}.) and elaborated in Algorithm~\ref{algo:cmf-eval-strategy}. 
The number and sites of faults to be injected depends on the selected fault campaign as discussed in Section~\ref{sec:attack-model}.

\begin{algorithm}[ht]
\footnotesize
\caption{\bf: Common Mode Failure (CMF) Evaluation}
\label{algo:cmf-eval-strategy}
\begin{algorithmic} [1]
    \Require $R$ Replicas 
    \Require CMF Compute Engine~\ref{fig:cmf-compute-engine}
    \State {\bf$SIM$} : Simulation Count
    \State {\bf$F$} : Failure rate of the circuit
    \State {\bf$Rel$} : Reliability (\%) of the circuit
    \State {$\mathbf{V}$}: Randomly generated 100,000 test vectors
    \State {\bf$CMF_{OUT}$}: Instantaneous CMF detector output 
    \State {\bf$K$}   : Failure Count
    \State $K\gets0$
    \For{$(i = 1 \to SIM)$}
        \State Inject faults w.r.t. the {\it Error Campaign}
        \State Apply $\mathbf{V}$ to the circuit with faults injected        
        \State Compute CMF~(\ref{eqn:cmf-det}) and save results
        \For{$(j = 1 \to \mathbf{V})$}     
            \State {\bf if} ($CMF_{j} \ne 0$) {\bf then} increment $K$
        \EndFor
    \EndFor
    \State $F=\left(\frac{K}{SIM}\right)$
    \State $Rel(\%)=(1 - F)\times100$
\end{algorithmic}
\end{algorithm}

\subsubsection{Diversity classes}
We evaluate (1) various \acronym configurations (variable $\alpha$ and $\beta$) compared to (2) TMR as a state-of-the-art replication method and (3) without replication as a baseline (No-TMR). The focus here is to create diverse N-bit adders.
We do not consider other CMAs due to space limitations and given that the N-bit adder is a good representative example.
For simulations and evaluations, the $16-bit$ Ripple-Carry adder (RCA) artifact is considered. However, the proposed approach can be extended to any value of $N$. 

A selected set of diverse CMA Adder implementations were generated with different diversity ratios are used across all experiments. The selected versions were enough to generate hundreds of permutations and millions of intrusion/fault injections.
The selected modules--functionally equivalent but structurally different--along with the gate count and failure probability ($P_f$) are shown in Table~\ref{table:diverse-modules}. They are also classified into three resilience levels which depends on the respective failure probabilities of the module. 
The selected modules have varying fault tolerance, area and logic level.
Area is measured in terms of the number of logic gates, while the logic depth shows the longest path between an input and an output.
$P_f$ denotes the fault coverage against \texttt{stuck-at-0} and \texttt{stuck-at-1} faults and is computed by performing simulations against a single fault using the HOPE fault simulator~\cite{Hope1996}. 
From Table~\ref{table:diverse-modules} it is also evident that, as the number of gates increase, $P_f$ (not CMF) decreases due to the logical masking of the modules.

\begin{table}
\centering
\caption{4-bit Full Adder (FA) modules ($M_i$)~\cite{afzaal2022evolutionary}.}
\label{table:diverse-modules}
\begin{tabular}{|c|c|c|c|c|}
\hline
\textbf{\begin{tabular}[c]{@{}c@{}}Resilience\\ Level\end{tabular}} & \textbf{\begin{tabular}[c]{@{}c@{}}Modules \\ ($M_i$)\end{tabular}} & \textbf{\# Gates} & \textbf{\begin{tabular}[c]{@{}c@{}}Logic \\ Levels\end{tabular}} & \textbf{\begin{tabular}[c]{@{}c@{}}Prob. Failure \\ ($P_f$)\end{tabular}} \\ \hline
\multirow{3}{*}{\textbf{Low}} & $M_1$ & 20 & 9 & 0.90 \\ \cline{2-5} 
 & $M_2$ & 24 & 9 & 0.85 \\ \cline{2-5} 
 & $M_3$ & 25 & 11 & 0.82 \\ \hline \hline
\multirow{3}{*}{\textbf{Medium}} & $M_4$ & 27 & 9 & 0.77 \\ \cline{2-5} 
 & $M_5$ & 28 & 9 & 0.76 \\ \cline{2-5} 
 & $M_6$ & 29 & 10 & 0.70 \\ \hline \hline
\multirow{4}{*}{\textbf{High}} & $M_7$ & 31 & 10 & 0.66 \\ \cline{2-5} 
 & $M_8$ & 32 & 10 & 0.64 \\ \cline{2-5} 
 & $M_9$ & 34 & 9 & 0.57 \\ \cline{2-5} 
 & $M_{10}$ & 37 & 10 & 0.56 \\ \hline
\end{tabular}
\end{table}

\subsubsection{Simulating Attacks}
\label{subsec:simulating-attacks}
To simulate Distribution faults, same faults are injected at the same location in the common modules among replicas. Evaluation is done by injecting faults across all possible locations in common modules. For example, consider Fig.~\ref{subfig:i4d4}, the common modules are $M1$ and $M4$ among the two replicas: faults will be injected at the same locations of $M1$ and $M4$ in these replicas. However, the total fault injection combinations will be the cross product of the number of gates in the respective common modules i.e., $|M1|\times|M4|$.

In the Zonal fault model, faults are injected at the common zones of the redundant replicas. The zone granularity is a module. In Fig.~\ref{fig:replicas-manifestation}, replicas are divided into four zones as each one of them has four modules. This approach emulates the faults that could be manifested due to external attacks on the same spatial zones of the replicas. 
$\sum_{zones(z)} \prod_{Replicas(i)} |M_{i,z}|$
Where $|M_{i,z}|$ denotes the total number of gates in a module of replica $i$ in zone $z$. The zonal fault injection campaign is more compute intensive in comparison to the distribution faults as it takes into account combinations of all possible locations in the replicas.
The distribution and zonal attacks are combined in the \textit{Compound attack} model by injecting faults in common modules among replicas and then iterating through zones for zonal fault injection.  \emph{It should be noted that we only report the detection of CMFs and expect that there is a correction mechanism in place for the rectification of faults if the replicas' outputs are different from each other.}

\begin{figure}[t]
    \centering
    \includegraphics[width=0.7\linewidth]{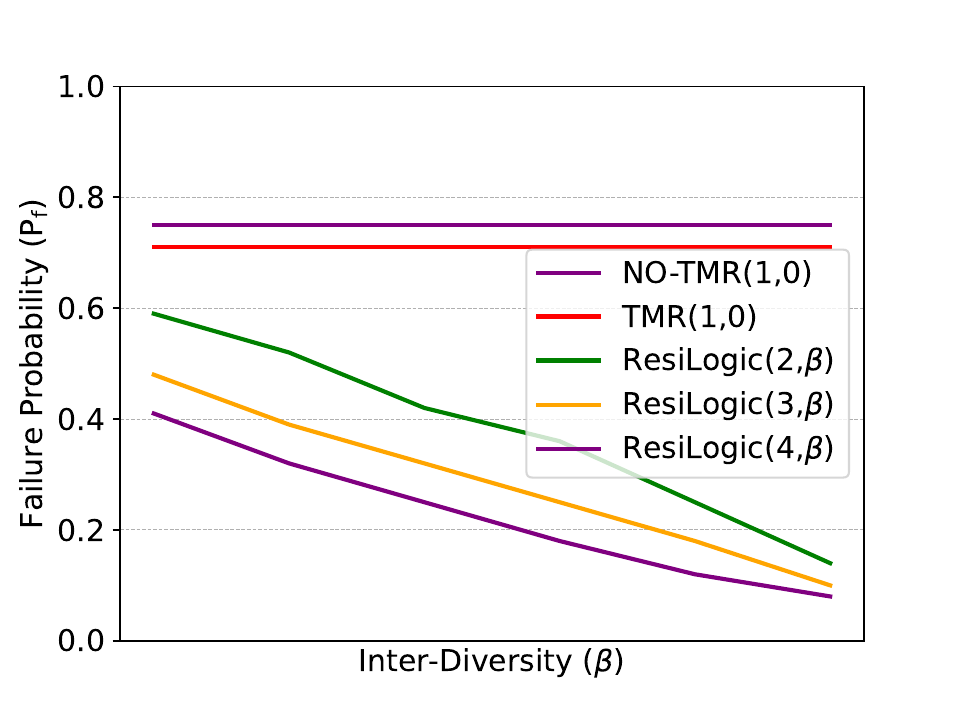}
    \caption{\acronym ($\alpha,\beta$) comparison with TMR and NO-TMR.}
    \label{fig:distribution-line-plot}
\end{figure}

\begin{figure*}[ht]   
      \begin{subfigure}{0.33\textwidth}
           \includegraphics[width=\textwidth]{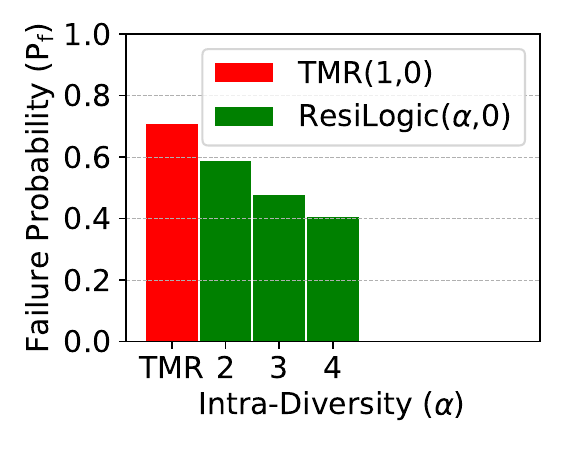}
            \caption{$\beta=0$}
            \label{fig:fixed-beta0}
    \end{subfigure}%
     \begin{subfigure}{0.33\textwidth}
           \includegraphics[width=\textwidth]{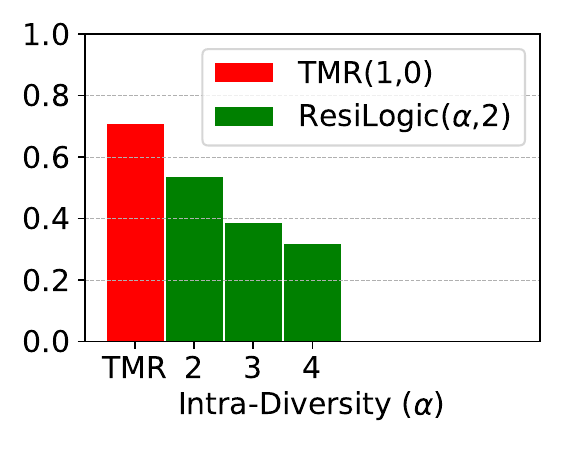}
            \caption{$\beta=2$}
            \label{fig:fixed-beta2}
    \end{subfigure}%
     \begin{subfigure}{0.33\textwidth}
           \includegraphics[width=\textwidth]{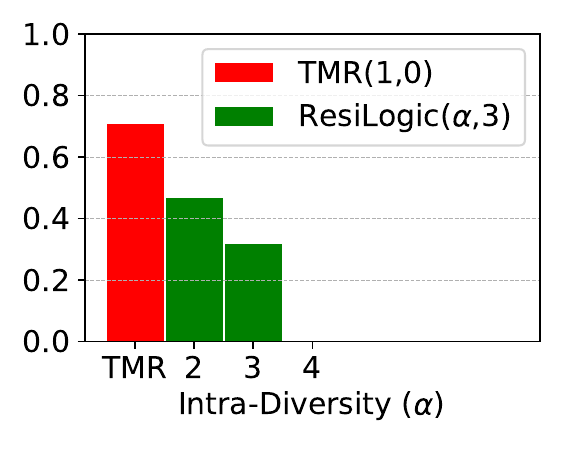}
            \caption{$\beta=3$}
            \label{fig:fixed-beta3}
    \end{subfigure}%
    
     \begin{subfigure}{0.33\textwidth}
           \includegraphics[width=\textwidth]{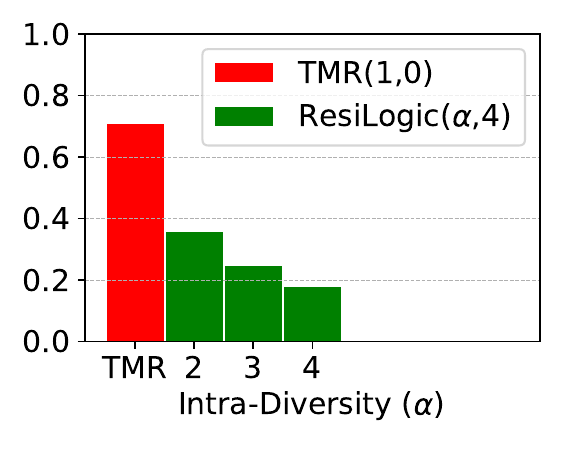}
            \caption{$\beta=4$}
            \label{fig:fixed-beta4}
    \end{subfigure}%
     \begin{subfigure}{0.33\textwidth}
           \includegraphics[width=\textwidth]{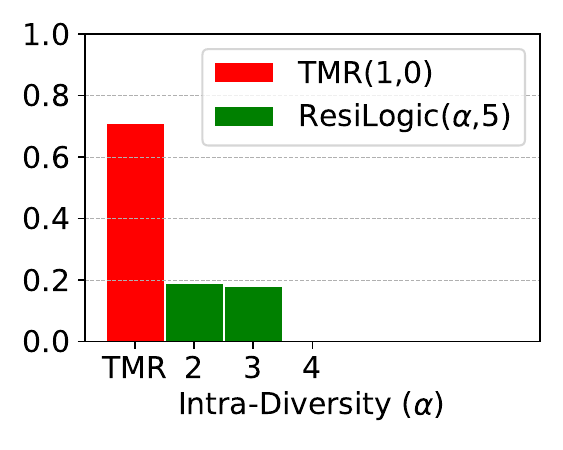}
            \caption{$\beta=5$}
            \label{fig:fixed-beta5}
    \end{subfigure}%
     \begin{subfigure}{0.33\textwidth}
           \includegraphics[width=\textwidth]{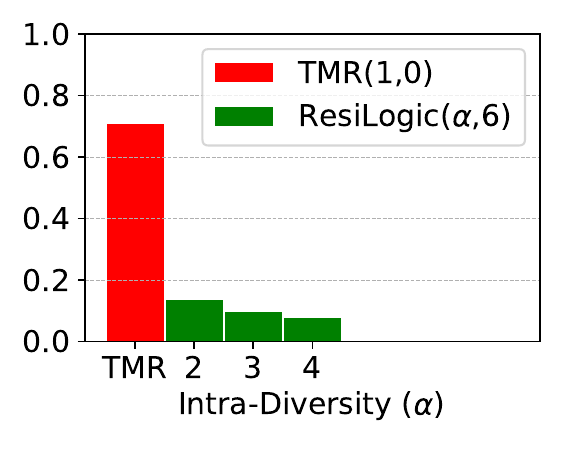}
            \caption{$\beta=6$}
            \label{fig:fixed-beta6}
    \end{subfigure}%
    \caption{\acronym's ($\alpha,\beta$) resilience to Distribution faults/attacks as intra-diversity ($\alpha$) vary.}
    \label{fig:dist-faults-beta}
\end{figure*}

\begin{figure*}[ht]   
    \begin{subfigure}{0.33\textwidth}
            \includegraphics[width=\textwidth]{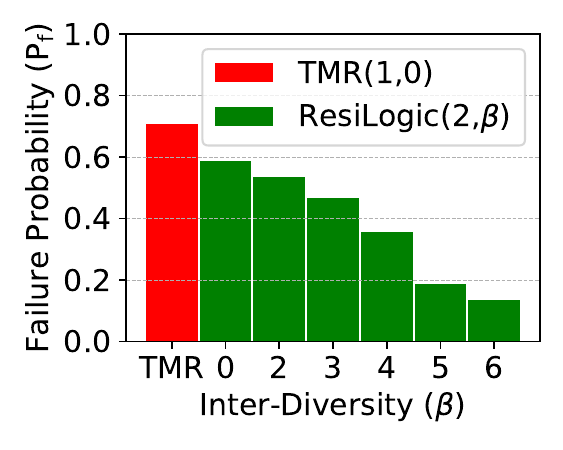}
            \caption{$\alpha=2$}
            \label{fig:fixed-alpha2}
    \end{subfigure} 
    \begin{subfigure}{0.33\textwidth}
            \includegraphics[width=\textwidth]{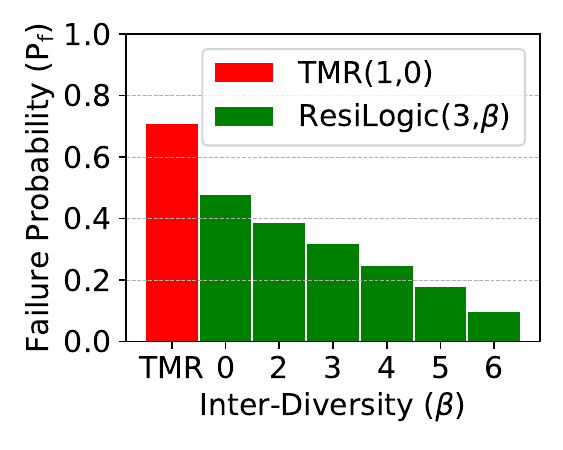}
            \caption{$\alpha=3$}
            \label{fig:fixed-alpha3}
    \end{subfigure}        
    \begin{subfigure}{0.33\textwidth}
           \includegraphics[width=\textwidth]{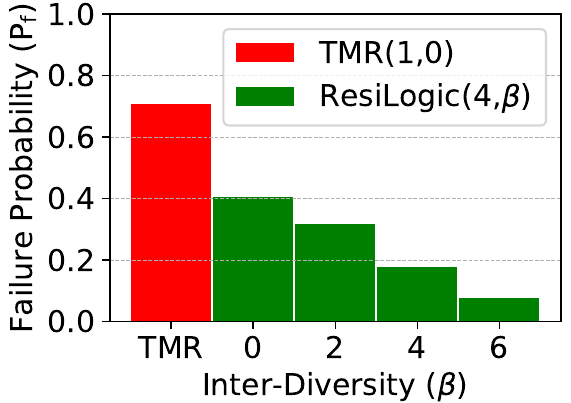}
            \caption{$\alpha=4$}
            \label{fig:fixed-alpha4}
    \end{subfigure}%
    \caption{\acronym's ($\alpha,\beta$) resilience to Distribution faults/attacks as inter-diversity ($\beta$) varies.}
        \label{fig:dist-faults-alpha}
\end{figure*}

\subsection{Resilience under Distribution Attack/Fault}
We first discuss the general patterns of failure probability for \acronym against TMR (without composition diversity) as state-of-the-art solutions and No-TMR (no replication) as baselines. We observe in Fig.~\ref{fig:distribution-line-plot} that all configurations of \acronym are significantly more resilient than the baselines. In particular, as more Inter-diversity is induced across replicas \acronym's resilience ranges between 0.6 and 0.1, while both TMR and No-TMR (whose lines are separated for visibility) exhibit around 0.7 resilience. This shows that TMR is not effective without diversity as it fails similar to No-TMR due to common mode failures. 

We then discuss Intra- and Inter-diversity aside, showing that both are important and complementary. We omit No-TMR for clarity, knowing that its resilience is similar to TMR across all Distribution attack results. Fig.~\ref{fig:dist-faults-beta} highlights the resilience for fixed $\beta$, while intra-diversity $\alpha$ varies between 2 and 4. The results consistently show that diversity by composition improves resilience of \acronym over TMR as $\alpha$ increases to 10\% with each diverse module added. The improvement is significant for higher $\beta$. In the best case (f) where $\beta=6$, i.e., different modules are used in different replicas, the failure probability is lower than 0.1. These results are consistent with the case where $\alpha$ is fixed, whereas the Intra-diversity $\beta$ varies in Fig.~\ref{fig:dist-faults-alpha}. With few diverse modules (a) $\alpha=2$ per replica, the resilience improves as long as more diverse modules are used in different replicas. The improvement is significant in the most diverse configuration $\alpha=4$. The interesting observation in varying $\alpha$ and $\beta$ is that the designer can tune the level of diversity based on the available versions, cost, and sensitivity of the application.

\subsection{Module placement and alignment across replicas}
\begin{figure}[t]
    \centering
    \includegraphics[width=0.8\linewidth]{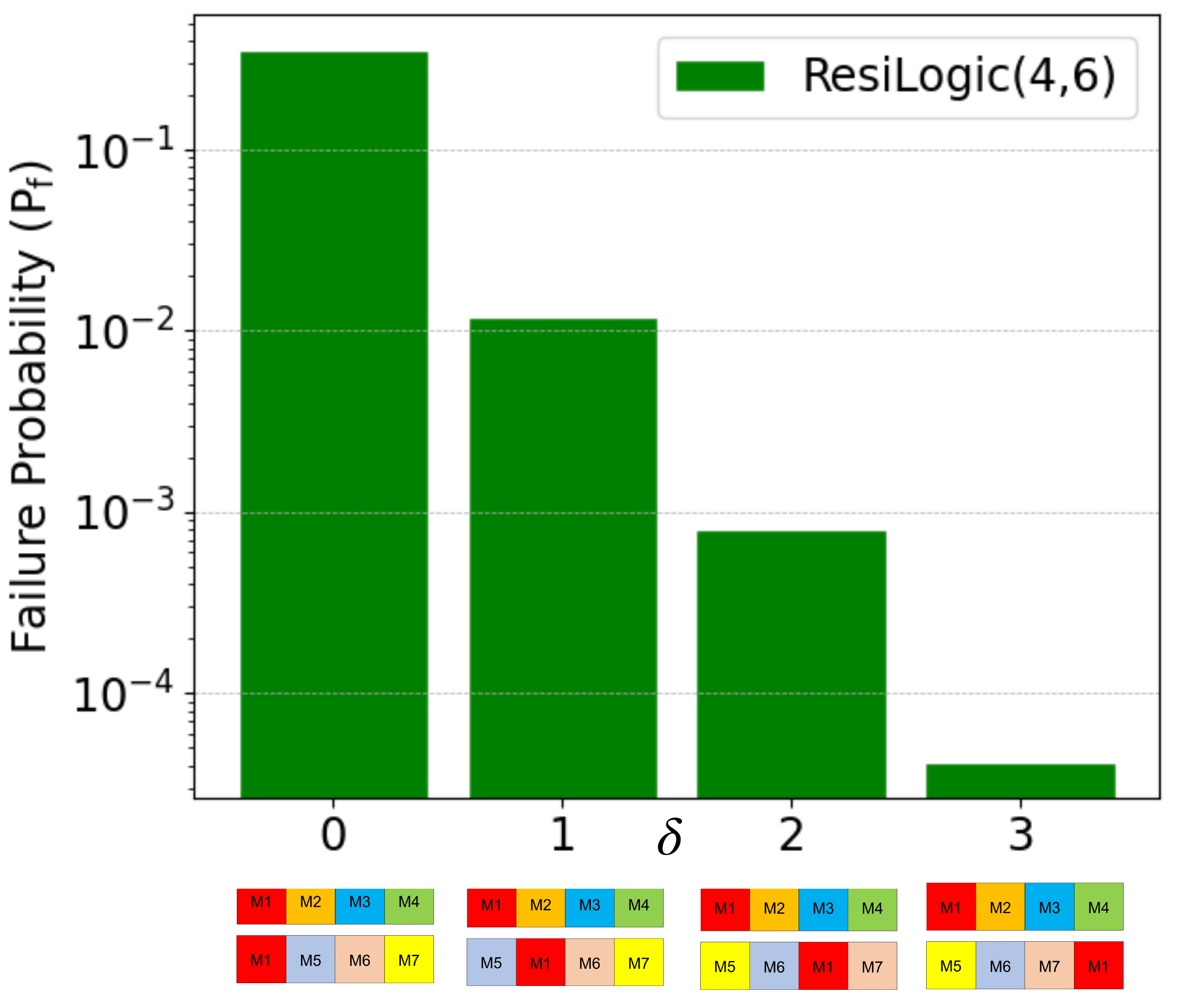}
    \caption{\acronym's resilience to Distribution faults/attacks, varying modules' placement in two replicas. Failure probability is lower when common modules are distant.}
    \label{fig:mod-position}
\end{figure}

An impactful property of composition in \acronym is the position of common modules across replicas that plays an important role in overall resilience. We explain this in Fig.~\ref{fig:mod-position}, where two replicas $R1$ and $R2$ are constructed in such a way that $R1=\{M1, M2, M3, M4\}$ and $R2=\{M5, M6, M7, M1\}$. $M1$ is the common module and the placement distance difference between the location of $M1$ in two replicas is $3$ i.e., $M1$ location in $R1$ and $R2$ is $0$ and $3$, respectively. From Fig.~\ref{fig:mod-position}, it can be observed that increasing alignment distance between a common module across two replicas improves resilience because varying common module location in replicas has a direct impact on fault controllability and observability conditions, resulting in reduced CMFs. 

\subsection{Resilience under Zonal Attacks}
We now evaluate \acronym under Zonal attacks. We pick two simulated scenarios: a horizontal attack (e.g., Electromagnetic field applied Fig.~\ref{fig:zonal-attacks}) which affects an entire replica and its modules, and a vertical attack that affects one module of each replica. Since No-TMR has no replication, the entire modules of the CMA are attacked. As expected, under the horizontal zonal attack an entire CMA replica is attacked in TMR and \acronym(4,8) (full diversity).  The impact of the attack is 100\% masked by the majority voting. The No-TMR case however fails drastically with failure probability close to 0.9. This reinforces the fact that TMR is useful in some cases even without modular diversity under Zonal attacks. The advantage is, however, not present in the case of vertical zonal attack, since all replicas in TMR and \acronym are hit by the attack, which cannot be masked by the voter. As expected, the failure probability in this case is only affected by the gate-level diversity of the modules used. 

\begin{figure}[t]
    \centering
    \includegraphics[width=0.8\linewidth]{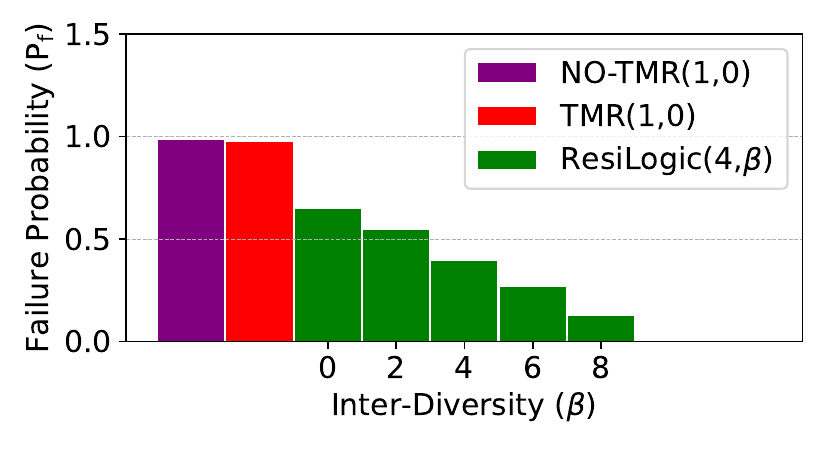}
    \caption{\acronym's ($\alpha,\beta$) resilience to Compound faults/attacks (simultaneous Distribution \& Zonal attacks).}
    \label{fig:compound-faults}
\end{figure}

\subsection{Resilience under Compound Attacks}
We finally consider the case of Compound attack where the Zonal and Distribution attacks are simultaneously applied. 
We only choose to evaluate with one \acronym (best) configuration due to space limitations (and since the trends are similar in others). 
We fix $\alpha=4$, while varying the value of $\beta$ in Fig~\ref{fig:compound-faults}. For the case of $\alpha=1,\beta=0$, all modules are victims of the Distribution attack which has a huge impact on TMR and No-TMR as we have seen in Fig.~\ref{fig:distribution-line-plot}. For the case where replicas are composed of diverse modules and there are no common modules among them, a module is picked at random for the injection of the Distribution attack. 
\acronym can sustain the Compound attack with down to 0.1 failure probability compared to TMR and No-TMR which fail most of the time.

\subsection{Impact on Area, Power and Delay}
To determine the impact of proposed technique on area, power and delay, the replicas are generated with specific resilience according to the classification levels mentioned in Table~\ref{table:diverse-modules}. The area, delay, and logic levels are calculated using the SIS synthesis tool~\cite{singh1992sis} for the MCNC library. The MCNC library is slightly modified to include only 2-input AND/OR, NAND/NOR and XOR/XNOR gates along with an Inverter and Buffer. This is done to make sure the library remains highly correlated with the designs mentioned in Table~\ref{table:diverse-modules}.
The area is computed by adding the fanins of all gates in a circuit. The dynamic power consumption of a digital circuit is determined for the MCNC library operating at a supply voltage of 5V and a frequency of 20 MHz. 

Algorithm~\ref{alg:proposed-framework} is applied by fixing the values of $\alpha$ and $\beta$ to $\{1,0\}$, $\{1,8\}$ and $\{4,8\}$ respectively. For the combination $\{4,8\}$ 100 CMAs were generated by considering all the modules from $(M_1...M_{10})$ in creating CMAs (Eqn.~\ref{eqn:cla}) as shown in Table~\ref{table:diverse-modules}. Table~\ref{table:area_power_delay} provides valuable insights into the trade-offs between area, power, and delay across different redundancy configurations. The results are averaged, as there are multiple circuits generated against each value of $\alpha$ and $\beta$. It can be observed that the area and power overheads are negligible across all profiles, while the delay is more impacted by the depth level of the graph. 
The case $\{1,8\}$ is a TMR but with different replicas and it can be observed that the area, power, delay and level steadily increase for each resilience level. This makes sense as higher resilience means more levels, which has a direct impact on other qualitative parameters. From Fig.~\ref{fig:compound-faults} we observed that the case of $\{4,8\}$ has the best resilience against the compound attacks, and it is also evident here in Table~\ref{table:area_power_delay} that the overheads are still close to the TMR. This is due to \acronym fine grained diversity mechanism.

\begin{table}
\centering
\caption{Qualitative profiles highlighting Area, Power and Delay for few diverse cases.}
\label{table:area_power_delay}
\begin{tabular}{|c|c|c|c|c|c|}
\hline
\textbf{$(\alpha,\beta)$} & \textbf{\begin{tabular}[c]{@{}c@{}}Modules\textquotesingle\\ Resilience\end{tabular}} & \textbf{\begin{tabular}[c]{@{}c@{}}Area\\ (Gates)\end{tabular}} & \textbf{\begin{tabular}[c]{@{}c@{}}Power\\ ($\mu W$)\end{tabular}} & \textbf{\begin{tabular}[c]{@{}c@{}}Delay\\ ($ns$)\end{tabular}} & \textbf{Levels} \\ \hline \hline
(1,0) & TMR & 1351 & 3035 & 65 & 41\\ \hline \hline
 & Low & 1347 & 3021 & 60 & 39 \\ \cline{2-6} 
 & Medium & 1352 & 3038 & 72 & 43 \\ \cline{2-6} 
\multirow{-3}{*}{(1,8)} & High & 1357 & 3051 & 74 & 46 \\ \hline \hline
(4,8) & $\forall M_i$ & 1350 & 3034 & 68 & 42 \\ \hline
\end{tabular}
\end{table}

%% file: conclusion.tex
\section{Conclusions}
\label{sec:conclusion}
Existing approaches to resilient chip design have mostly been studied under benign fault models where no malicious actors exist~\cite{inoue2009gmr}. \acronym addresses that gap, motivated by recent digital sovereignty concerns~\cite{us_chip_act-2022, eu_chip_act-2023}.
Productive chip businesses focus on design, while making use of off-the-shelf tools and libraries from different vendors after which fabrication is outsourced. If vendors and fabs are not trusted, the design can be vulnerable to several attacks, among them three we introduced: Distribution, Zonal, and Compound.
To mitigate these attacks, we introduced the notion of \textit{Diversity by Composability} that enables the building of resilient circuits out of smaller diverse modules at a low cost, thus reducing the typical replication factor overhead.
We also provided a technique to use E-Graphs to generate diverse replicas at gate-level. 
The quality of generated diverse replicas is directly influenced by the \texttt{rewrite} rules. 
Together with state of the art gate-level diversity and modular redundancy, we show that resilience to the aforementioned attacks is significantly improved up to a factor of five, and the designer retains some control on diversity.